\documentclass[12pt,a4paper]{article}
\usepackage[margin=2cm]{geometry}
\usepackage[margin=5pt,labelfont=bf]{caption}
\usepackage{graphicx,amsmath,amssymb,natbib,setspace,
  xspace,color,subcaption,appendix,authblk}
\definecolor{darkblue}{rgb}{0.0,0.0,0.6}
\usepackage[colorlinks=true,breaklinks=true,linkcolor=darkblue,
urlcolor=darkblue,anchorcolor=darkblue,citecolor=darkblue]{hyperref}

\newcommand{\diff}[2]{\dfrac{\text{d}{#1}}{\text{d}{#2}}}

\newcommand{\cotwo}{\ensuremath{\text{CO}_2}}
\newcommand{\fcotwo}{\ensuremath{f_{\text{CO}_2}}}
\newcommand{\ecotwo}{\ensuremath{E_{\text{CO}_2}}}
\newcommand{\gcotwo}{\ensuremath{\mathcal{G}_{\text{CO}_2}}}
\newcommand{\taumin}{\ensuremath{\tau_{\text{min}}}}
\newcommand{\taumax}{\ensuremath{\tau_{\text{max}}}}
\newcommand{\taumean}{\ensuremath{\tau_{\text{mean}}}}
\newcommand{\Sdot}{\ensuremath{\dot{S}}}

\title{Variations in mid-ocean ridge \cotwo{} emissions driven by
  glacial cycles}

\author[$\dag$]{Jonathan M.A.~Burley\thanks{jonathan.burley@earth.ox.ac.uk}} 
\author[$\dag$]{Richard F.~Katz} \affil[$\dag$]{\small Department of
  Earth Sciences, University of Oxford, UK}

\begin{document}
\maketitle

\begin{abstract}
  The geological record documents links between glacial cycles and
  volcanic productivity, both subaerially and, tentatively, at
  mid-ocean ridges. Sea-level-driven pressure changes could also
  affect chemical properties of mid-ocean ridge volcanism.  We
  consider how changing sea-level could alter the \cotwo{} emissions
  rate from mid-ocean ridges on both the segment and global scale.  We
  develop a simplified transport model for a highly incompatible trace
  element moving through a homogenous mantle; variations in the
  concentration and the emission rate of the element are the result of
  changes in the depth of first silicate melting.  The model predicts
  an average global mid-ocean ridge \cotwo{} emissions rate of
  $53$~Mt/yr or $91$~Mt/yr for an average source mantle \cotwo{}
  concentration of 125 or 215 ppm by weight, in line with other
  estimates.  We show that falling sea level would cause an increase
  in ridge \cotwo{} emissions about 100~kyrs after the
  causative sea level change. The lag and amplitude of the response
  are sensitive to mantle permeability and plate spreading rate.  For
  a reconstructed sea-level time series of the past million years, we
  predict variations of up to $12\%$ in global mid-ocean ridge
  \cotwo{} emissions.
\end{abstract}

\section{Introduction}
Glacial cycles transfer $\sim 5 \times 10^{19}$~kg of water between
the oceans and ice sheets \citep{Tushingham:1991tr}, leading to
accumulation and ablation of kilometres of ice on the continents and
sea-level change of $\sim100$~m. In Iceland, for example, the pressure
change associated with melting of the ice sheet since the last glacial
maximum had well-documented consequences for the volcanic activity
\citep{Sigvaldason:1992fv,Jull:1996bs,Maclennan:2002ba} and lava
geochemistry \citep{Maclennan:2002ba}.  More broadly, continental
volcanism in both the northern and southern hemispheres shows
increased activity associated with the last deglaciation
\citep{Gardeweg:1998iu,Jellinek:2004hb,Huybers:2009hn}.
\cite{Huybers:2009hn} and \cite{Lund:2011jd} hypothesised that the
pressure variations caused by changing sea level during glacial cycles
would affect mid-ocean ridge (MOR) volcanism.  \cite{Crowley:2014vy}
documented variations of bathymetry near the Australian-Antarctic
ridge that are possible evidence of such glacial effects.

A simple argument shows that variations in crustal thickness and
sea-floor relief should be expected to result from sea-level
variation. The melting rate of a parcel of mantle beneath a MOR is
proportional to its depressurisation rate.  As the parcel upwells, it
depressurises due to the decreasing height of rock above it.  The rate
of change of pressure due to upwelling is the gravitational
acceleration $g$ times the mantle density times the upwelling rate
($\sim 3$~cm/yr). Sea-level variation can modify this depressurisation
rate: the pressure change due to varying sea level is the product of
$g$, water density and the rate of change of sea level (up to 100~m in
10~kyr or 1~cm/yr).  Water is about one third the density of the
mantle and sea-level changes can be equivalent to one third the mantle
upwelling rate, implying that sea-level changes can modify
depressurisation rates, and hence melting rates, by up to $\pm10$\%.
\cite{Crowley:2014vy} apply a paleo-sea-level reconstruction to a
simulation of MOR melting and melt transport, with melting rates
varying according to the variable pressure exerted by sea level. This
leads to varying melt flux at the ridge, predicting variations in
crustal thickness consistent with bathymetric observations of
sea-floor relief.

Given this evidence for glacial cycles affecting MOR melt production,
it is reasonable to consider if, as in Iceland
\citep{Maclennan:2002ba}, the chemistry of the lavas is also affected.
To investigate this we develop a model of the transport of a highly
incompatible element from the asthenosphere through the melting region
to the MOR.  Highly incompatible elements partition strongly into the
melt, rather than remaining in the residual solid.  Approximating this
as complete incompatibility creates useful simplifications in
modelling.  For instance, a completely incompatible element's path
through the melting region is entirely determined by the motion of the
melt, without any need to consider that element in the solid or
partitioning between phases.  Furthermore, for small perturbations to
the melting rate (such as those caused by sea-level change) the
mass-flow rate of the element through the MOR is constant (see
appendix~\ref{sec:inst-co2-resp}).

In a simple model of melting beneath a mid-ocean ridge, a parcel of
mantle upwells adiabatically beneath the ridge axis, cooling slightly
due to its expansion. The pressure-dependent solidus temperature of
that parcel decreases as it ascends; at some depth (or, equivalently,
pressure), the temperature of the parcel is equal to its solidus
temperature.  This depth, thought to be around $60$~km, is called the
depth of first silicate melting.  With further upwelling, the parcel's
temperature exceeds the solidus and it partially melts. As soon as the
first increment of melt is produced, 100\% of the completely
incompatible element that was locally present in the solid mantle is
transferred to the melt.  Because the mantle is permeable and the melt
is less dense than the residue, the melt ascends faster than the
solid, segregating from its source.  Melt segregation and transport of
incompatible elements thus begins at a pressure-dependent depth. More
specifically, melt segregation begins at a fixed pressure, but the
depth corresponding to this pressure can change. We assume that there
is no isostatic rebound associated with sea-level change and hence
mantle upwelling is constant.  Variations in sea level will therefore
cause the depth of first silicate melting (and initiation of melt
segregation) to rise and fall.  The rate at which mantle crosses this
boundary and delivers its content of incompatible elements to the
melting region is the mantle upwelling rate minus the rate of upward
motion of the boundary.  So in this model, variations in the depth of
first silicate melting cause variations in the flux of an incompatible
element. As sea level falls, the depth of first melting increases,
upwelling mantle crosses into the melting region faster, and the flux
of incompatible element increases; the reverse is true for sea-level
rise.  Any perturbation to the melting rate within the melting region
does not alter the mass of the incompatible element in the melt, it
only dilutes (or concentrates) the element. Variations in
melt-transport rate associated with melting perturbations are a
secondary effect and are not considered in detail here (though see
appendix~\ref{sec:inst-co2-resp}).

A more nuanced view of melting may disagree with this simple story in
some of the details, especially with the inclusion of volatile
elements that are present in small concentrations in the
mantle. Experimental evidence suggests that \cotwo{}-rich melt forms
at $\sim250$~km depth \citep{Dasgupta:2013us} and has a low viscosity
that rises sharply with silica content at shallower depth
\citep{Kono:2014ja}. If such carbon-rich melts can segregate from the
solid mantle it would complicate the role of the transition to
silicate melting at around $60$~km.  However, it remains an open
question whether oxygen fugacity allows such melts to form and, if
they do form, whether such tiny melt fractions can segregate from the
solid mantle.  \cite{Dasgupta:2013us} suggests carbonatite melt
fractions reach $\sim0.03$~wt\% deep in the mantle below ridges, which
is at the lowest limit of carbonatite melt interconnectivity of
$0.03$--$0.07$~wt\% in $\sim0.05$~mm olivine grains
\citep{Minarik:1995vq}.  The additional presence of water might
increase the melt fraction to $0.06$--$0.1$~wt\%
\citep{Dasgupta:2013us} by 150~km depth, but the threshold for
interconnectivity of such melts is not known.  In our calculations, we
assume that these melt fractions do not segregate from the solid
mantle until the onset of volatile-free peridotite melting at
$\sim60$~km raises the melt fraction, creating an interconnected,
permeable network of pores.

Among the highly incompatible elements, we focus on carbon despite its
active role in the thermodynamics of melt production because
variations in \cotwo{} emissions from the solid Earth are potentially
important to understanding past variation of the climate.  The solid
Earth contains $10^{10}$--$10^{11}$~Mt carbon \citep{Dasgupta:2010fw},
orders of magnitude more than the atmosphere ($0.6\times10^{6}$~MtC
\citep{Solomon:2012uy}) and the oceans ($4\times10^{7}$~MtC
\citep{Solomon:2012uy}).  Solid-Earth carbon emissions from MORs are
estimated as $\sim25$~MtC/yr \citep{Coltice:2004jl, Cartigny:2008dz,
  Marty:1998vo} and from arc volcanoes $\sim20$~MtC/yr
\citep{Coltice:2004jl}.  As the largest reservoir, the solid Earth's
carbon budget is known to control atmospheric carbon on multi-Myr
timescales; geological ages show a correlation between volcanic
activity and atmospheric \cotwo{} concentration \citep{Budyko:1987}.
Hence there is evidence for both MOR volcanism being affected by
glacial cycles and for the effect of volcanic \cotwo{} emissions on
atmospheric \cotwo{} concentration.  While we focus on \cotwo{} in our
model, the same theory applies equally to other highly incompatible
elements such as U, Th, Nb, Ba, and Rb.

The model is developed under the guiding principle that it should be
simple enough that the connections between the assumptions and the
outputs are readily traceable.  The full model is assembled from
independent, decoupled parts that capture the key physics with minimal
complexity.  Mantle flow is modelled by the passively-driven
corner-flow solution \citep{batchelor67:Book,Spiegelman:1987wd};
lithospheric temperature structure and thickness is computed with a
half-space cooling model \citep{Geodynamics:turcotte}.  To quantify
melt generation and transport we use one-dimensional compaction
columns \citep{Ribe:1985ua, Hewitt:2010il} that are based on
conservation of energy, mass, and momentum at steady state in a
homogenous mantle.  The outline of the melting region is given by a
parameterised solidus \citep{Katz:2008gd}.  A focusing width is
applied such that melt focused to the ridge produces a maximum crustal
thickness of $7$~km.  A detailed discussion of the assumptions made in
deriving the model is presented below.

To summarise the results, the model predicts that a section of MOR
spreading at $3$~cm/yr half-rate will see a change in the rate of
efflux of highly incompatible elements (\textit{e.g.}, \cotwo) of
$\sim 8$\% for a linear sea-level change of $100$~m in $10$~kyrs.  For
reconstructed sea level data and the present distribution of plate
spreading rates, the model predicts global MOR emissions to deviate
from the mean by up to $\pm 6$\%.  These results are sensitive to the
permeability of the mantle, which is a primary control on the rate of
melt transport.  There are good constraints on how permeability scales
with porosity, but its absolute value at a reference porosity (1\%
here) is uncertain. We consider a broad range of values that includes
the most extreme estimates.

Section~\ref{sec:the-model} details the model used to predict \cotwo{}
emissions for a section of mid-ocean ridge; parameter values are
stated in table~\ref{Tab:FixedParameters}.  The behaviour of the model
is demonstrated for simple scenarios of sea-level variation in
sections~\ref{sec:case-1:-baseline}--\ref{sec:case-3:-sawtooth}; the
model is applied to the global MOR system under a reconstructed
sea-level history in sections \ref{sec:reconstr-pleist-sea} and
\ref{sec:the-global-model}.  The results are discussed in
section~\ref{sec:discussion} and the key conclusions stated in
section~\ref{sec:conclusion}.

\section{The Model} \label{sec:the-model}
\begin{figure}[ht]
  \centering
  \includegraphics[width=8cm]{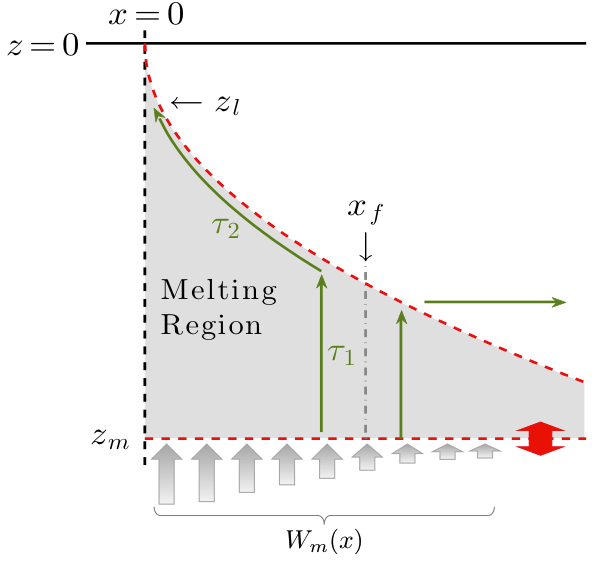}
  \caption{Sketch of the melting region. Two example melt streamlines are
    shown in green. Mantle upwelling rate into the base of the melting region
    $W_m$ is represented by grey arrows, the size of which shows the
    decreasing magnitude of $W_m$ with distance from the ridge axis.  The
    red arrow indicates the variability of $z_m$ with respect to time,
    expressed in equation~\eqref{eq:MeltingDepthDiff} (all other
    boundaries are steady state).  The melt streamlines on either side
    of the maximum focusing distance $x_f$ show melt flow to the ridge
    or frozen into the lithosphere.}
  \label{fig:MeltRegion_Sketch}
\end{figure}

Our goal is to develop a method to compute the \cotwo{} emission rate
\ecotwo{} (mass per unit time per unit length of ridge) from a segment
of ridge.  To achieve this we require a model of \cotwo{} flux into
the melting region and also of its transport to the ridge.  We
approximate the behaviour of \cotwo{} as perfectly incompatible, and
hence there is no exchange of \cotwo{} between phases during melt
transport.  The rate of ridge emission of \cotwo{} can then be
quantified by integrating the mass flux into the base of the melting
region $f_{CO2}$ (mass/area/time) and using the travel time from the
base of the melting region to the ridge. This is formulated as
\begin{align}
  \label{eq:E_CO2}
  \ecotwo(t) &= 2\int_{0}^{x_f}f_{CO2}(t_s,x,U_0) \text{d}x \;\;,
\end{align}
where $x$ is the horizontal distance from the ridge axis, $x_f$ is the
maximum distance over which melt is focused to the ridge axis, $U_0$
is the half-spreading rate, and $t$ is time.  A parcel of melt
arriving at the ridge axis at time $t$ was produced by mantle that
crossed into the melting region at time $t_s$.  The travel time of
melt from the base of the melting region to the ridge is represented
as $\tau$, which varies with lateral distance $x$ from the ridge axis.
Hence the source time is $t_s(x)=t-\tau(x)$. The factor of two in
eqn.~\eqref{eq:E_CO2} arises from the symmetry of the melting region
across the ridge axis. A sketch of half the melting region is shown in
figure~\ref{fig:MeltRegion_Sketch}.

Equation~\eqref{eq:E_CO2} requires an expression for $f_{CO2}$. This
is a product of the rate at which mantle material crosses the depth of
first silicate melting $z_m$ and the \cotwo{} concentration in that
material.  For generality, we consider an expression that allows for
volatile-enriched insipient partial melting beneath the depth of first
silicate melting, though we will later exclude this scenario from
consideration. Thus, \fcotwo{} is written as
\begin{equation}
  \label{eq:LONG_CO2_flux_rate}
  f_{\cotwo}(t_s,x,U_0)  =\left(W_m(x,U_0)-\diff{z_{m}(t_s)}{t}\right) 
  \left(1- \phi \right)\mathcal{C}_{\cotwo}^{\textrm{solid}} + 
  \left(w_m-\diff{z_{m}(t_s)}{t}\right) 
  \phi \; \mathcal{C}_{\cotwo}^{\textrm{melt}} \;\;,
\end{equation}
where $W_m(x,U_0)$ is the upwelling rate of the mantle, $w_m$ is the
upwelling rate of insipient melt, $\mathcal{C}_{\cotwo}$ is a mass
concentration of \cotwo{}, and $\phi$ is the volume fraction of melt;
all of these are evaluated at depth $z_m$ and distance $x$ from the
ridge axis.  The first term in parentheses on the right-hand side of
the equation is the rate at which mantle crosses into the melting
region.  The concentrations and melt fraction may be considered
steady-state, constant values as long as
$w_m \ge W_m \ge \text{max}\left(\dot{z}_m(t)\right)$.  This ensures
that the rate at which material crosses the depth of first silicate
melting is always positive or zero.  Even the fastest sea-level
changes on record, meltwater pulses during the last deglaciation,
satisfy these conditions for most MORs: 20~m sea level change in
500~years gives $\dot{z}_m=1.3$~cm/yr.  However, meltwater pulse
events are not resolved in the reconstructed sea-level series that we
consider in this paper \citep{siddall10}, so the conditions
$w_m \ge W_m \ge \text{max}\left(\dot{z}_m(t)\right)$ are true with
only occasional exceptions for the slowest spreading ridges.  With
these conditions satisfied and assuming that either $\phi(z_m)=0$ or
that $w_m = W_m$, equation~\eqref{eq:LONG_CO2_flux_rate} can be
simplified to
\begin{equation}
  f_{\cotwo}(t_s,x,U_0) =\left(W_m(x,U_0)-\diff{z_{m}(t_s)}{t} \right) \mathcal{C}_{\cotwo}.
  \label{eq:CO2_flux_rate}
\end{equation}
Here we should interpret $\mathcal{C}_{\cotwo}$ as the mass
concentration of \cotwo{} in the solid mantle plus co-moving insipient
melt, if it is present. This model could be modified to accommodate
more complicated situations, but this is not given further
consideration at present.

The flux of \cotwo{} in equation~\eqref{eq:CO2_flux_rate} depends on
the solid mantle upwelling rate at the depth of first silicate
melting.  Approximating this as passive (plate-driven) flow of
isoviscous rock, mantle upwelling is given by the corner flow solution
\citep{batchelor67:Book, Spiegelman:1987wd}.  The vertical component
of this solution, evaluated at $z=z_m$, can be written as
\begin{equation}
  W_m(x,U_0) =  \frac{2U_0}{\pi-2\alpha_c-\sin{2\alpha_c}} 
  \left( \frac{1}{1+\frac{x^2}{z_m^2}} - \sin^2(\alpha_c) \right) \;,
  \label{eq:W_x}
\end{equation}
where the lithosphere is represented as a wedge with angle
$\alpha_{c}$ to the horizontal.  We follow \cite{Spiegelman:1987wd} in
computing the wedge angle to approximately match the plate thickness
at a specified distance from the axis.  We use
$\alpha_c = \tan^{-1}(z_l/x_w)$, such that the wedge intersects the
upper boundary of the melting region $z_l$ at the maximum width of
melting region $x_w$ (see appendix~\ref{sec:deriv-melt-width} for
details).  This definition of the wedge angle ensures that the
lithospheric wedge does not overlap with the melting region, and that
the upwelling rate is small but non-zero at the extreme width of the
melting region ($0 \leq W_m(x_w) \ll U_0$).

An expression for the depth of first melting $z_m$ is needed in
equations~\eqref{eq:CO2_flux_rate} and \eqref{eq:W_x}.  Taking
decompression as the only influence on local mantle temperature prior
to melting, we model the depth of first melting as the intersection
between an adiabatic temperature profile and the solidus temperature
profile.  We approximate both profiles as linear with respect to depth
(details in appendix~\ref{sec:deriv-depth-melt}) to obtain
\begin{equation}
  z_m  =  - \Bigg(\frac{\widetilde{T}-T_{S_{\textrm{ref}}}}{\gamma \rho g  
    - \frac{\alpha g \widetilde{T}}{c} }\Bigg) + \frac{\rho_w}{\rho} S\;\;,
  \label{eq:MeltingDepth}
\end{equation}
where $\widetilde{T}$ is the mantle potential temperature,
$T_{S_{\textrm{ref}}}$ is the solidus temperature at reference mantle
composition and surface pressure, $\gamma$ is the Clausius-Clapeyron
slope for the mantle, $\rho$ is the mantle density, $\alpha$ is the
coefficient of thermal expansion, $c$ is the specific heat capacity,
$S$ is the sea-level deviation from a long-term mean, and $\rho_w$ is
the density of water. The first term in
equation~\eqref{eq:MeltingDepth} is the dry peridotite melting depth
of $\sim60$~km, and the second term is the shift in melting depth due
to sea-level.  The only time dependent variable in
equation~\eqref{eq:MeltingDepth} is the sea level $S$, so
differentiating gives
\begin{equation}
  \diff{z_{m}}{t} = \frac{\rho_w}{\rho}\diff{S}{t}.
  \label{eq:MeltingDepthDiff}
\end{equation}
These equations state that silicate melting begins at a fixed
pressure, but the depth corresponding to this pressure varies as sea
level rises and falls.

The source time $t_s$ used in equations~\eqref{eq:E_CO2} and
\eqref{eq:CO2_flux_rate} depends on the travel time of melt to the
ridge from any point $x$ along the base of the melting region.  For
simplicity, we consider melt flow as following a vertical path from
the base to the top of the melting region, then following a high
porosity channel along the impermeable top of the melting region to
the ridge axis, as illustrated by streamlines $\tau_1$ and $\tau_2$ in
figure~\ref{fig:MeltRegion_Sketch}.  This is a reasonable
approximation of numerically modelled streamlines for homogenous
mantle \citep[\textit{e.g.},][]{Katz:2008gd}, where buoyancy forces
drive vertical fluid flow in the majority of the melting region, with
the compaction pressure only becoming large enough to deflect melt
flow from the vertical within a few kilometres of melt-impermeable
boundaries \citep{Sparks:1991th}.  We consider flow along the high
porosity channel as instantaneous, motivated by the high flow rates
expected there, as compared to vertical flow rates in the rest of the
melting region \citep{Katz:2008gd}.  To compute the travel time from
the base of the melting region to the base of the lithosphere, we use
a 1D compaction column from \cite{Hewitt:2010il}.  This model assumes
Darcy flow and thermodynamic equilibrium for a two-component,
homogenous mantle with a constant Clausius-Clapeyron slope.  Following
the reduced model of \cite{Crowley:2014vy}, we assume that small
variation in melting rates due to sea-level change do not
significantly affect melt velocity or travel time.  Furthermore, in
computing $\tau$ we take $z_m$ as constant, because changes in $z_m$
due to sea level are only tens of metres, changing $\tau$ by
$\ll 1 \%$.  This gives a travel time
\begin{equation}
  \tau(W_m,z_l,z_m) = \phi_0\left(\frac{\eta_f}{K_0\Delta\rho g}\right)^{\frac{1}{n}}
  \left(\frac{\rho}{\Pi W_m}\right)^{1-\frac{1}{n}} 
  \left[n \left(z_l-z_m\right)^{\frac{1}{n}}\right] \;,
  \label{eq:tau}
\end{equation}
where $\eta_f$ is the mantle melt viscosity, $K_0$ is the permeability
of the mantle at the $1\%$ reference porosity $\phi_0$, $\Delta\rho$
is the density difference between the solid and melt, $n$ is the
porosity exponent in the permeability relation
($n \approx 2.7 \sim 3$, \cite{Miller:2014gp}), $z_l(x,U_0)$ is the
depth of the upper boundary of the melting region, and $\Pi$ is the
adiabatic melt productivity (kg of melt produced per m$^3$ per m
upwelling).  This productivity is given by \cite{Hewitt:2010il} as a
ratio of thermodynamic parameters, and we retain the same parameter
values here.

To close the model we need an expression for the upper boundary of the
melting region $z_l(x)$.  This boundary is located where the local
temperature equals the solidus temperature, which is controlled by the
thickness of the conductively cooled boundary layer that forms the
lithosphere.  \cite{Crowley:2014vy} also included the effects of
adiabatic decompression and latent heat removal, but this leads to an
expression for $x_l(z)$ that cannot be inverted.  However, as shown in
appendix~\ref{sec:deriv-upper-bound}, the depth of this melting
boundary is approximated by an isotherm of the half-space cooling
model, which can be expressed as $z_l(x)$.  This isotherm is hotter
than the low-pressure solidus, but the temperature difference
compensates for the change in solidus temperature with respect to
depth.  We use
\begin{equation}
  \label{eq:Lithospheric_thickness_1}
  z_l(x,U_0)= 2\sqrt{\frac{\kappa x}{U_0}}\; \text{erfc}^{-1}
  \left(\frac{\widetilde{T}-T_l}{\widetilde{T}-T_0}\right)\;,
\end{equation}
where $T_l$ is the temperature of the upper boundary of melting
region, assumed to be constant.  $T_0$ is the temperature of the ocean
floor, and $\kappa$ is the thermal diffusivity of the mantle.  Note
that although we have neglected adiabatic decompression and latent
heat of melting in equation~\eqref{eq:Lithospheric_thickness_1}, these
are accounted for in the melting calculations.

\begin{table}[h!]
  \begin{center}
    \begin{tabular}{crll}
      \hline     
      Parameter		&	Value	&	&	Parameter description					\\
      \hline
      $c$			&	$1200$	&	J/kg/K&	Specific heat				\\
      $\mathcal{C}_{\cotwo}$		&	$0.375$	& 	kg/m$^3$		& CO$_2$ mass per m$^3$ of mantle (derived from 125ppm by weight)		\\
      $g$			&	$10$	&	m/s$^2$		&	Gravitational acceleration		\\
      $K_0$		&	$10^{- 13}$--$10^{- 11}$	&	m$^2$		&	Permeability at 1\% porosity		\\
      $K$			&	$K_0 (\phi / \phi_0)^n$	&	m$^2$		&	Permeability		\\
      $L$			&	$4 \times 10^{5}$	&      J/kg	&	Latent heat of the mantle				\\
      $L_{\text{MOR}}$	&	$61\,000$	& 	km   &	Total length of the mid-ocean ridge system	\\ 
      $n$			&	$3$	&	-      &      Porosity exponent in the permeability relation	\\
      $S$			&	[various]	&	m		&	Sea level, expressed as deviation from long term mean \\
      $\dot{S}$		&	[various]	&	cm/yr		&	Rate of change of sea level with respect to time \\
      $\widetilde{T}$	&	$1648$	&	K		&	Potential temperature of upwelling mantle	\\ 
      $T_{0}$		&	$273$	&	K		&	Ocean floor temperature	\\
      $T_{\textrm{s}_{\textrm{ref}}}$	&	$1554$	&	K		&	Reference solidus temperature	\\
      $U_0$		&	$\leq 8$	& 	cm/yr		& Plate half-spreading rate \\
      $x_f$			&	$10$--$70$	& 	km
      & Width of region, at $z_m$, from which melt is focused to the ridge\\
      $x_w$		&	$10$--$350$	& 	km		& Width of melting region \\
      $z_m$			&	$60$	& 	km   &	Depth of first melting	\\
      $\alpha$		&	$3\times 10^{- 5}$	&
      K$^{-1}$	&	Thermal expansion	coefficient for the mantle		\\
      $\gamma$	&	$60\times 10^{- 9}$	&	K/Pa	&
      Clausius-Clapeyron slope of the mantle		\\
      $\eta_f$		&	$1$		&	Pa$\,$s	&	Mantle melt viscosity			\\
      $\kappa$		&	$10^{- 6}$	&	m$^2$/s	&	Thermal diffusivity	\\
      $\Pi$			&	$0.01$	&	kg/m$^{4}$
      &	Adiabatic melt productivity, kg of melt per m$^{3}$ of mantle \\ &&& per metre of upwelling			\\
      $\rho$		&	$3 300$	& 	kg/m$^3$		& Mantle density					\\
      $\rho_c$		&	$2 900$	& 	kg/m$^3$ &	Oceanic crust (mean) density					\\
      $\rho_w$		&	$1 000$	& 	kg/m$^3$ &	Freshwater density	\\
      $\Delta \rho$	&	$500$	& 	kg/m$^3$ &	Density difference between liquid and solid mantle		\\
      $\phi_0$		&	$0.01$	&	-		&      Reference porosity (volume fraction)			\\
      \hline
    \end{tabular}
  \end{center}
  \caption{Parameter values for calculations.}
  \label{Tab:FixedParameters}
\end{table}

\section{Results}
\label{sec:results}

Below we demonstrate the behaviour of the model by a series of
examples, then consider global \cotwo{} emissions for reconstructed
sea level.  We begin by modelling a unit length of mid-ocean ridge in
the absence of sea-level change.  This defines the baseline state of
the model that we compare against when applying sea-level forcing.
The first dynamic example is a single, linear sea-level change, which
illustrates key characteristics of the model and emphasises the
importance of the melt travel time. The emissions curve for this
example approximates the Green's function for the model: the response
to a step-change in sea level.  We next consider simple, periodic sea
level curves, demonstrating the system's response to an oscillatory
forcing as a function of the frequency of that forcing.  We then
compute the model's response to a reconstructed Pleistocene sea-level
record.

Building on these calculations for a single section of mid-ocean
ridge, we calculate predictions for a global model, created by summing
sections with appropriate spreading rates over the full MOR system.
We demonstrate this composite model by considering the global
emissions response to a single sea-level change and then to the
reconstructed Pleistocene sea-level curve.

\subsection{Constant Sea Level \& Baseline Emissions}
\label{sec:case-1:-baseline}

\begin{figure}[h!]
  \centering
  \includegraphics[width=9cm]{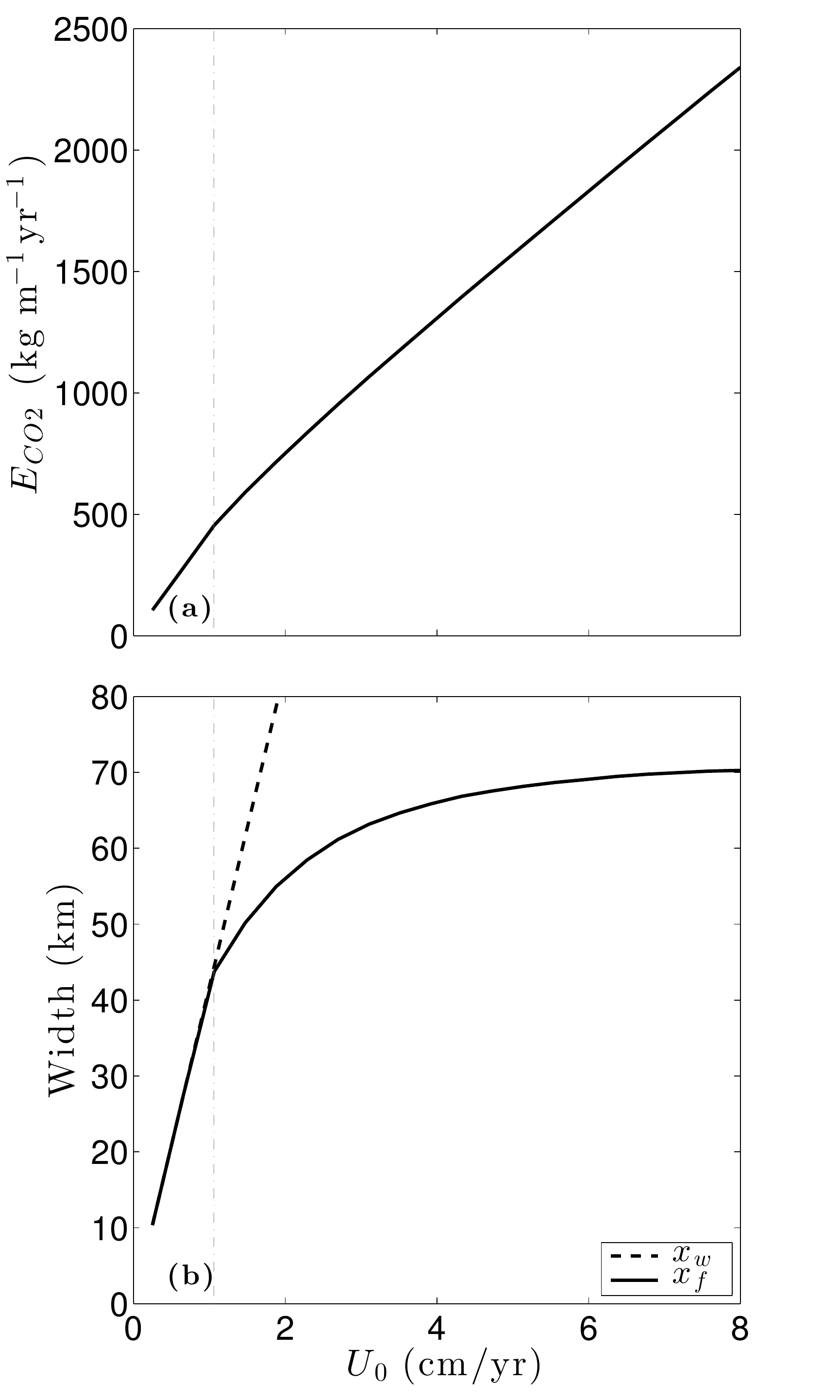}
  \caption{\textbf{(a)} Ridge \cotwo{} emissions for constant sea level, and
    \textbf{(b)} focusing width $x_f$ and full width $x_w$ of the melting
    region; all shown for varying half-spreading rate. Focusing width
    is equal to the width of the melting region when crustal thickness
    is less than $7$~km ($U_0 \leq 1$~cm/yr), and is otherwise limited
    such that crustal thickness does not exceed $7$~km.  The switch
    between these behaviours is marked by the grey dotted line at
    $1$~cm/yr.  \ecotwo{} is computed with \cotwo{} concentration in
    the mantle of 125~ppmw.}
  \label{fig:Results_Case1}
\end{figure}

Baseline emissions are defined as the steady-state emission rate of
\cotwo{} from the MOR for constant sea level.  This represents the
background state, which is disturbed after sea-level
change. Figure~\ref{fig:Results_Case1} shows baseline emissions as a
function of plate spreading rate and demonstrates that, in the absence
of changing sea level, \cotwo{} emissions per metre of ridge are
approximately proportional to the half-spreading rate $U_0$.  A faster
spreading rate drives faster mantle upwelling, bringing more \cotwo{}
into the melting region.  Faster spreading rates also increase the
width of the melting region (see appendix~\ref{sec:deriv-melt-width}),
leading to more melt and more \cotwo{} being focused to the ridge
axis.  However, not all melt produced is focused to the MOR; at some
lateral distance, melts are frozen back into the lithosphere, rather
than travelling to the ridge axis \citep[e.g.,][]{Katz:2008gd}.  We
incorporate this detail by enforcing a maximum focusing width $x_f$
such that melt focused to the MOR will produce crust $\leq 7$~km thick
in steady state, consistent with observations
\citep{White:1992ei,Bown:1994fh}. Crustal thickness is calculated as
the volume flow rate of melt to the ridge axis (kg~m$^{-3}$ per metre
along MOR axis) times the density ratio of basaltic melt to oceanic
crust, divided by the half-spreading rate. At half-rates
$U_0 \geq 1$~cm/yr, the focusing width is smaller than the width of
the melting region. With the imposition of this limit on melt
focusing, there is a slight change in the slope of the baseline
emission curve at $U_0=1$~cm/yr (fig.~\ref{fig:Results_Case1}).

For the range of $U_0$ on Earth of up to 8~cm/yr,
figure~\ref{fig:Results_Case1}(a) shows baseline emissions of up to
2300~kg~m$^{-1}$~yr$^{-1}$.  This is calculated using 0.38~kg \cotwo{}
per m$^3$ of upwelling mantle (125~ppm \cotwo{} by weight, hereafter
ppmw), based on an average of MOR source mantle of 50-200 ppmw
\citep{Dasgupta:2013un,Cartigny:2008dz,Saal:2002vl,Marty:1998vo,Salters:2004ct}.
These source mantle \cotwo{} concentrations are inferred by four
methods: \textit{(i)} \cite{Cartigny:2008dz} start with Nb
concentration in MORBs, use Nb/C ratios, and assume an average degree
of melting to calculate a \cotwo{} concentration in the source
mantle. \textit{(ii)} \cite{Saal:2002vl} use MORB melt inclusions to
measure \cotwo{} concentrations in the erupting mantle immediately
prior to degassing, then assume an average degree of melting to
calculate a \cotwo{} concentration in the source
mantle. \textit{(iii)} \cite{Marty:1998vo} use $^3$He concentrations
in the ocean to infer a $^3$He efflux from MORs. Then they apply a
He/C ratio to estimate carbon efflux from MORs, which is matched to a
\cotwo{} concentration in the erupting mantle using a (completely
degassed) crustal formation rate of 21~km$^3$/yr. \cotwo{} in the
source mantle is then calculated by assuming an average degree of
melting to generate MORB.  \textit{(iv)} \cite{Salters:2004ct} start
from major elements in the mantle and use a chain of element ratios to
derive a \cotwo{} concentration in the depleted mantle.  For all these
approaches, \cotwo{} concentration in the mantle is derived using
several assumptions and has large uncertainties.  The global MOR
efflux of \cotwo{} is calculated using fewer assumptions and has less
uncertainty.  An alternative constraint on mantle \cotwo{} content is
to match the global baseline emissions of the model to the estimated
global MOR efflux by choosing a concentration of \cotwo{} in the
source mantle of 215~ppmw (section~\ref{sec:the-global-model}).  This
is effectively substituting our average degree of melting into other's
calculations, and gives a result slightly higher than prior estimates
of $\mathcal{C}_{\cotwo}$.  For simplicity, we use a generally
accepted 125~ppmw $\mathcal{C}_{\cotwo}$ throughout this paper, but
restate key results for the 215~ppmw value.  This changes \cotwo{}
emissions by a factor of $1.7$, as emissions scale linearly with
$\mathcal{C}_{\cotwo}$ (see eqn.\eqref{eq:CO2_flux_rate}).

Subsequent sections present scenarios of changing sea-level.  In these
sections, plots depict emissions in terms of percentage difference
from baseline emissions for the appropriate half-spreading rate; these
percentage values apply to all highly incompatible elements and for
any $\mathcal{C}_{\cotwo}$.  Percentage results can be converted to
\cotwo{} mass via figure~\ref{fig:Results_Case1}.  The deviations from
the baseline emission rate are a consequence of non-zero $\dot{z}_m$
in equation~\eqref{eq:CO2_flux_rate}. By definition, total emissions
are equal to this deviation summed with baseline emissions.

\subsection{Single Sudden Change in Sea Level}
\label{sec:case-2:-linear}
\begin{figure}[h!]
  \centering
  \includegraphics[width=12cm]{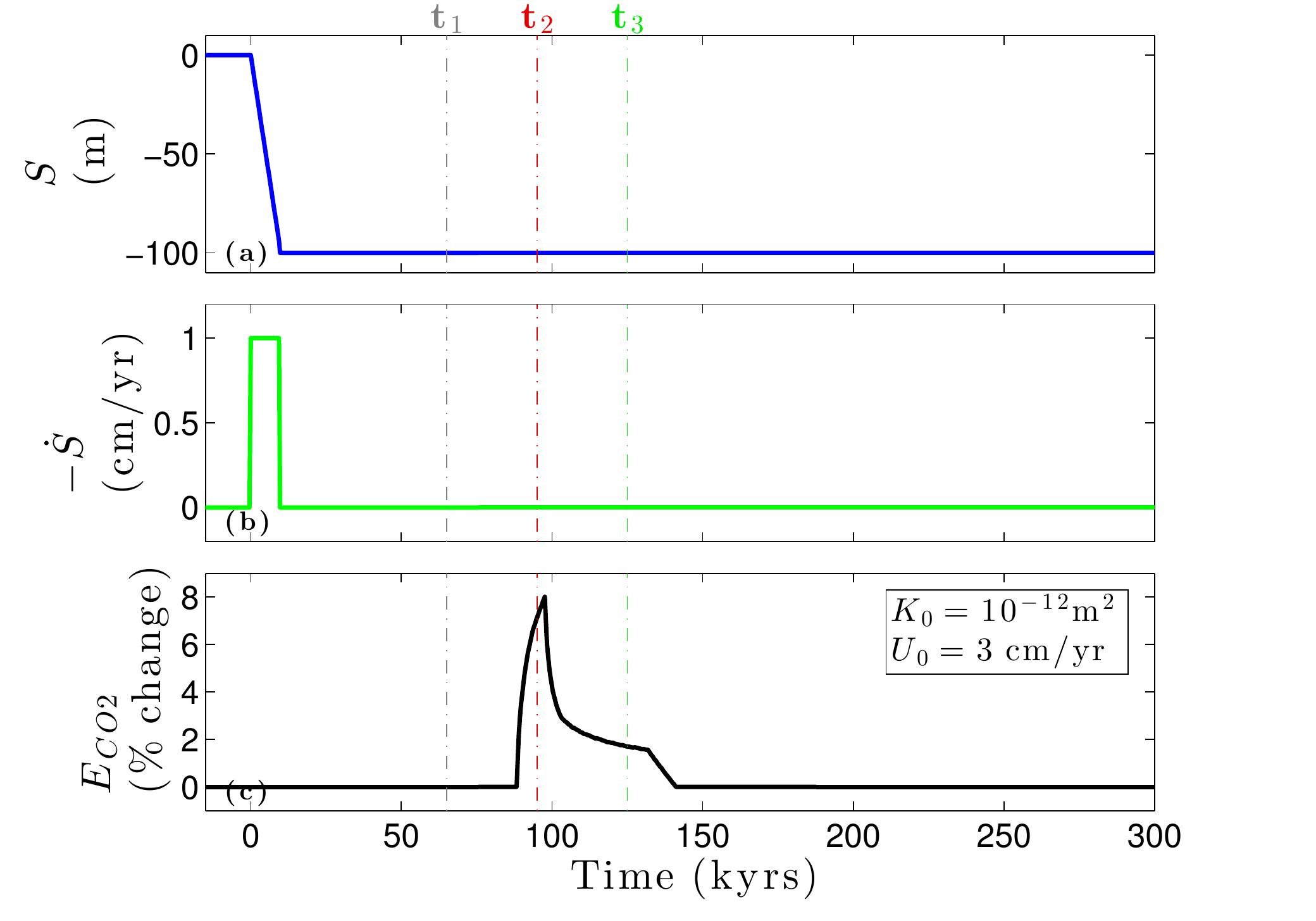}
  \caption{Single step change in sea level. Plots show
    \textbf{(a)} sea-level change, \textbf{(b)} the negative rate of
    sea-level change, and \textbf{(c)} \cotwo{} emissions from a
    section of mid-ocean ridge, measured as percentage change from
    baseline emissions.  The plate half-spreading rate and the
    permeability constant at $1\%$ porosity are $U_0$ and $K_0$,
    respectively.  The three times marked on the plots correspond to
    (a--i,ii,iii) in figure~\ref{fig:Results_Case2_Tau}. Negative
    $\dot{S}$ is plotted so that peaks in $\dot{S}$ and consequent
    peaks in \ecotwo{} point in the same direction.}
  \label{fig:Results_Case2}
\end{figure}

\begin{figure}[h!]
  \centering
  \includegraphics[width=17cm]{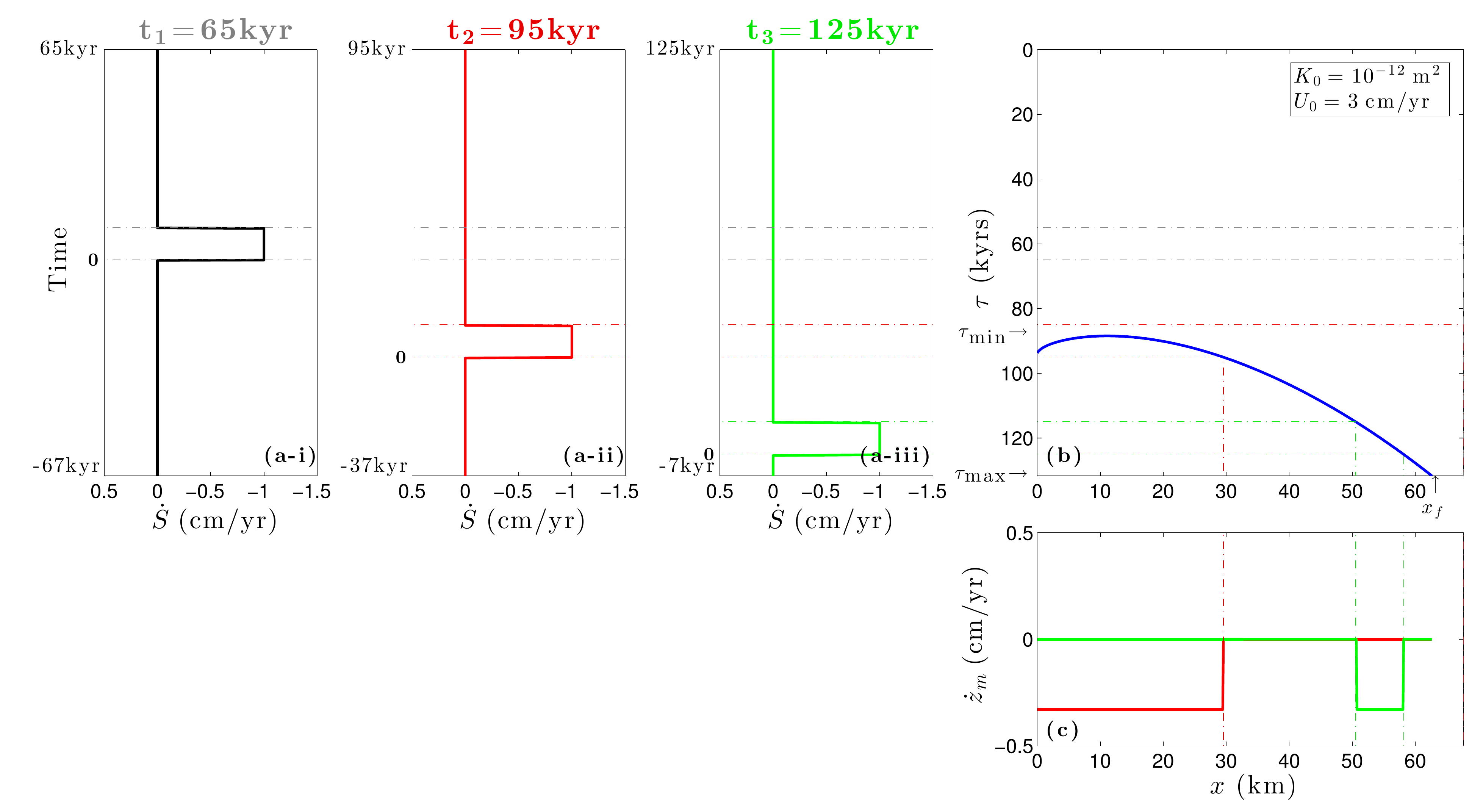}
  \caption{The effect of a linear change in sea level on ridge
    \cotwo{} emissions. The three points in time shown in black, red,
    and green show the state of the system $65$, $95$, and $125$~kyrs
    after the pulse began. Plots show: \textbf{(a--i,ii,iii)} rate of
    sea-level change from present into the past, for the three points
    in time marked in figure~\ref{fig:Results_Case2}. \textbf{(b)}
    Travel time of melt from the base of the melting region to the
    ridge, increasing downwards.  Travel time at $x=0$ is slightly
    greater than \taumin{} because the sharp increase in lithospheric
    thickness within a few km of the ridge increases the column
    height; the effect of this increase in height exceeds the effect
    of faster on-axis upwelling. \textbf{(c)} Rate-of-change of the
    depth of first melting that acted on the melt currently arriving
    at the ridge axis, when that melt began to segregate.  Dashed
    lines mark how the travel time converts $\dot{S}(t)$ to
    $\dot{z}_m(x)$. The integral of a coloured line from $0$ to $x_f$
    in panel~(c) is directly proportional to \ecotwo{} at the
    corresponding time in figure~\ref{fig:Results_Case2}(c).  This is
    expressed by the integral of $\dot{z}_m$ with respect to $x$ in
    equation \eqref{eq:CO2_flux_rate}.  A video animating this plot
    and \ecotwo{} over time is included in the online supplementary
    materials.}
  \label{fig:Results_Case2_Tau}
\end{figure}

Imposition of a sudden sea-level change exposes the behaviour of the
model.  We use a steep, linear ramp, which gives a box function in
$\dot{S}$. Figure~\ref{fig:Results_Case2} shows the predicted MOR
\cotwo{} emission rate \ecotwo{} resulting from this sea level
forcing. \cotwo{} emissions remain at the baseline level for
$\sim$90~kyrs after the change in sea level and then there is a sharp
rise in \ecotwo{}, representing an 8\% increase in emission rate.
This delayed response is due to the travel time of \cotwo{} from the
base of the melting region to the ridge. After the 8\% peak, \ecotwo{}
falls sharply to about 3\%, followed by a slow decay until 130~kyrs,
after which there is a linear drop back to the baseline level over the
duration of the box-pulse in $\dot{S}$. The origin of this emissions
pattern in figure~\ref{fig:Results_Case2}(c) is explored in
figure~\ref{fig:Results_Case2_Tau}.

Figure~\ref{fig:Results_Case2_Tau} demonstrates how the shape of the
emission response curve in figure~\ref{fig:Results_Case2}(c) is a
consequence of the travel time $\tau$ and its variation with respect
to distance $x$ from the ridge axis.  This travel time, shown in
figure~\ref{fig:Results_Case2_Tau}(c), ranges between \taumin{} for
melts originating in the mantle beneath the ridge axis and \taumax{}
for distal melts that are focused laterally.  The mixture of melts
that arrives at the ridge at time $t$ contains \cotwo{} that was
transported in melts that initiated along the base of the melting
region at all $x<x_f$.  These deep melts formed and began to segregate
from their source at times in the past $t-\tau(x)$.  The \cotwo{}
content of the segregated melt is different to the baseline case
according to the $\Sdot$ value at $t-\tau (x)$.  Therefore, we can
calculate the deviation from the baseline by considering the rate of
sea-level change (alternatively, $\dot{z}_m$) acting on the melting
region when each element of melt first segregated.  This is
represented in figure~\ref{fig:Results_Case2_Tau}(c), and the integral
of this plot with respect to $x$ is directly proportional to \ecotwo{}
in figure~\ref{fig:Results_Case2}.

The \ecotwo{} response is clarified by again considering a sharp,
linear drop of sea level that occurs over the interval $0\le t \le 10$
kyrs.  The drop in sea level causes the depth of the base of the
melting region to increase, importing additional \cotwo{} into the
melting regime. In panels (a-i), (a-ii), and (a-iii) of
fig.~\ref{fig:Results_Case2_Tau} we see the box pulse receding into
the past as time progresses over $t_1<t_2<t_3$
(figure~\ref{fig:Results_Case2}c shows \ecotwo{} at these times). The
start and end times of the pulse are projected onto $\tau(x)$ in panel
(b). At $t_1$, the projection lines do not intersect $\tau(x)$,
meaning that the \cotwo{} perturbation generated by $\Sdot$ has not
yet reached the ridge axis. Therefore the emissions curve in
figure~\ref{fig:Results_Case2}(c) remains at the baseline level.  At
$t_2$, the projection lines span $\taumin$; the shallow slope of
$\tau(x)$ near $\taumin$ means that \cotwo{} from a broad (30~km)
region affected by the \Sdot{} pulse is arriving at the ridge axis at
$t_2$.  This causes the spike in emissions shown in
figure~\ref{fig:Results_Case2}(c). At $t_3$, the interval of sea-level
change has receded far into the past.  The only \cotwo{} perturbation
in melts arriving at the ridge is in distal melts from the base of the
melting region at 50 to 58~km off-axis.  The narrowness of this band
translates to a reduced (but non-zero) value of \ecotwo{} at $t_3$ in
fig.~\ref{fig:Results_Case2}(c).  As the sea-level drop recedes
further into the past, the emission rate drops to zero because the
very distal melts (from $x>x_f$) are not focused to the ridge axis.
This process is animated in a video in the online supplementary
materials.

In the limit of vanishing duration of sea-level change, the pulse of
$\Sdot$ approximates to a Dirac delta function.  Hence the \ecotwo{}
response shown in figure~\ref{fig:Results_Case2}(c) is an
approximation of the Green's function.  Conceptually, the emissions
response for any sea-level time series could be approximated by
convolution with a Green's function approximation like
figure~\ref{fig:Results_Case2}(c).

From this discussion, it is clear that changes to $\tau{}(x)$ will
alter how MOR emissions respond to changing sea-level.
Equation~\eqref{eq:tau} for travel time shows that plate
half-spreading rate $U_0$ and permeability constant $K_0$ are the key
parameters affecting the melt travel-time $\tau$.  Higher values of
either lead to shorter travel times, giving a higher peak in \ecotwo{}
over a shorter time period, with this peak occurring sooner after the
causative sea-level change.  This is explored in more detail in the
following section.

\subsection{Oscillating Sea Level}\label{sec:case-3:-sawtooth}
We now consider oscillatory sea level and discuss the concepts of
lag and admittance.

\begin{figure}[Ht]
  \centering
  \includegraphics[width=17cm]{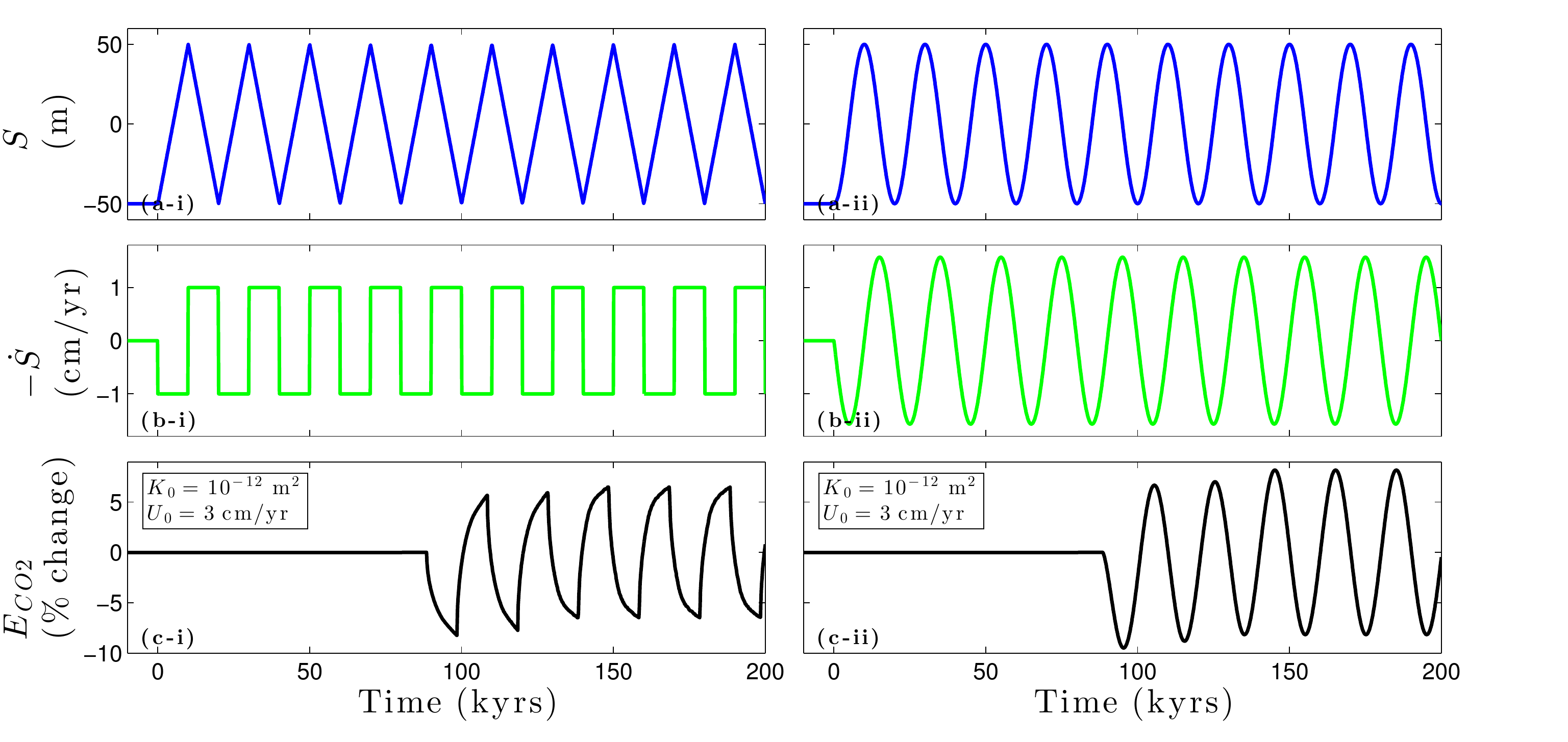}
  \caption{Sawtooth and sine waves in sea level starting at
    $t=0$. Plots show \textbf{(a)} sea-level, \textbf{(b)} negative
    rate of sea-level change, \textbf{(c)} \cotwo{} emissions from a
    section of mid-ocean ridge.  Considering the left column, each
    box-pulse in \Sdot{} produces an emissions peak/trough as in
    figure~\ref{fig:Results_Case2}(c) with interference where they
    overlap, giving the net result seen in (c-i). The steady upward
    shift in the \ecotwo{}-peaks until $\sim$150~kyr is due to this
    overlap. If the first box-pulse in $\dot{S}$ had been in the
    opposite direction, \ecotwo{} peaks in (c) would be mirrored
    across $\ecotwo{}=0$.}
  \label{fig:Results_Case3_50kyr}
\end{figure}

Figure~\ref{fig:Results_Case3_50kyr} shows a pair of oscillating sea
level scenarios and their predicted \ecotwo{} variation.  The left
column (i) shows a time series of alternating box-pulses in $\dot{S}$;
the right column (ii) shows a sinusoidal sea-level variation.
Figure~\ref{fig:Results_Case3_50kyr}(c-i) has an oscillating series of
peaks in \ecotwo{} resulting from a series of $\dot{S}$ box-pulses in
figure~\ref{fig:Results_Case3_50kyr}(b-i).  This $\dot{S}$ series is
equivalent to summing single box-pulses from
figure~\ref{fig:Results_Case2}(b), with suitable offset and amplitude.
Similarly, the \cotwo{} emissions can be represented as a sum of
offset emissions spikes from single, linear changes in sea level. The
train of emissions peaks in figure~\ref{fig:Results_Case3_50kyr}(c-i)
and (c-ii) stabilises in amplitude after $t \approx \taumax$; this
transient represents the spin-up time of the model, associated with
the tail of excess emissions shown in fig.~\ref{fig:Results_Case2}.

\begin{figure}[h!]
  \centering
  \includegraphics[width=17cm]{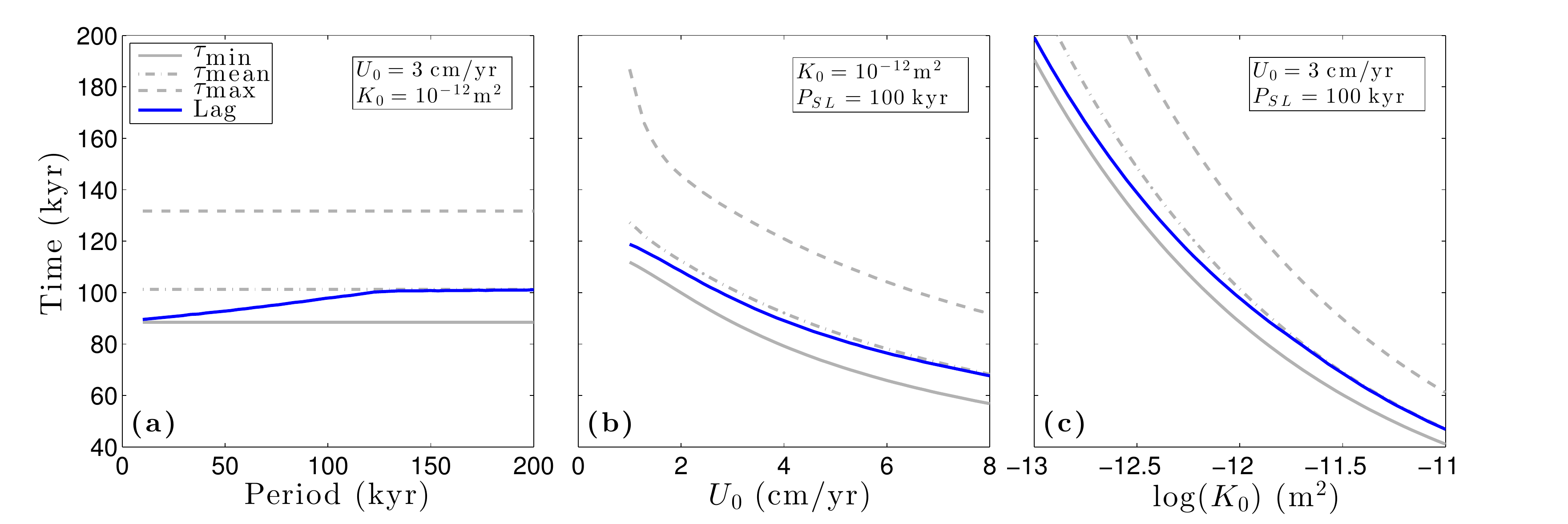}
  \caption{ Lag and travel time for varying \textbf{(a)} sinusoidal
    sea-level period, \textbf{(b)} half-spreading rate, and
    \textbf{(c)} permeability constant.  The lag is calculated as the
    time between a peak in $-\Sdot$ and the corresponding peak in
    \ecotwo{} for sinusoidal sea level.}
  \label{fig:Results_Case4_Delay}
\end{figure}

Figure~\ref{fig:Results_Case3_50kyr}(ii) shows sinusoidal sea level
and provides the context to define the lag metric.  Lag $\mathcal{L}$
is the time between a peak in $\dot{S}$ and the corresponding peak in
emissions. Because the time interval around \taumin{}~kyrs before $t$
has the largest influence on \ecotwo{} at $t$, the \ecotwo{} signal
should lag \Sdot{} by about \taumin{}. However, the lag will not be
exactly \taumin{}, as the influence of \Sdot{} on \ecotwo{} is felt up
to \taumax{} years after the change in sea level.  Thus the exact
value of lag will be slightly greater than \taumin{} and we expect
this difference to depend on the period of sea-level oscillation
relative to $\taumax-\taumin$.  In particular, when the period
approaches and exceeds $2(\taumax-\taumin)$, the lag becomes equal to
the mean melt-transport time \taumean{}.
Figure~\ref{fig:Results_Case4_Delay} shows lag, \taumin{}, $\taumean$,
and \taumax{} for varying half-spreading rate, permeability constant,
and sinusoidal sea-level period.  We note that
$\taumin \le \mathcal{L} \le \taumean \ll \taumax$ and therefore we assume
$\mathcal{L}\approx \taumin$ with a small, systematic error.

\begin{figure}[h!]
  \centering
  \includegraphics[width=12cm]{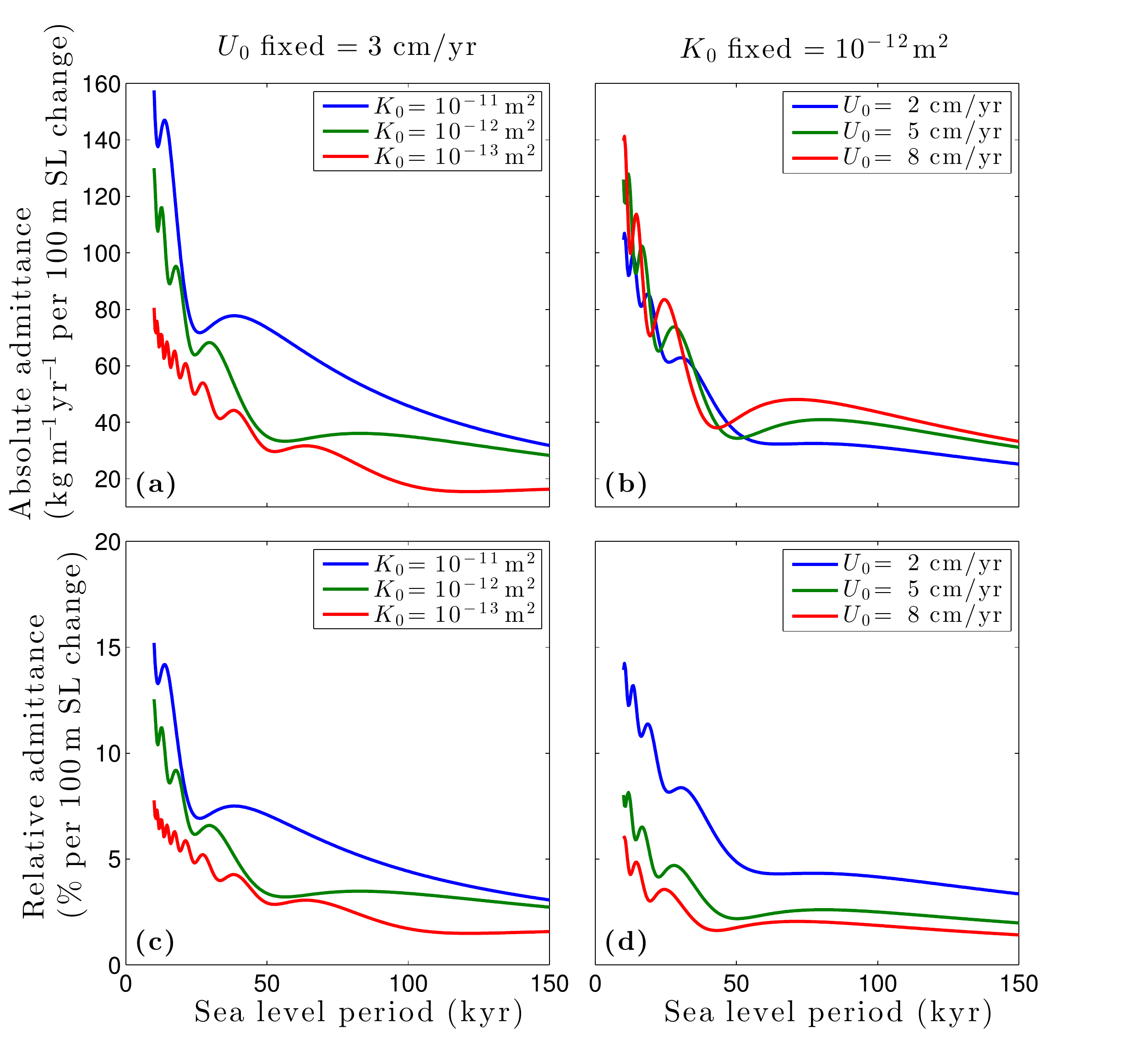}
  \caption{Absolute and relative admittance for varying
    half-spreading rate, permeability, and sinusoidal sea-level
    period.  Plots hold either $U_0$ or $K_0$ fixed.  The magnitude of
    sea-level change is constant for all periods. }
  \label{fig:Results_Case4_Admittance}
\end{figure}

Sinusoidal variation of sea level also provides a context in which
compute admittance. Admittance is the ratio of the response amplitude
to the forcing amplitude as a function of the sinusoidal forcing
period.  We define two versions of admittance: absolute admittance,
with units of kilograms \cotwo{} per metre of ridge per year per 100~m
of sea-level change, and relative admittance, with units of percentage
change from baseline emission rate per 100~m of sea-level change.  The
latter is the absolute admittance divided by the baseline emission
rate.  Figure~\ref{fig:Results_Case4_Admittance} shows absolute and
relative admittance and how they vary with changing sea-level period,
half-spreading rate, and permeability constant.  We discuss both the
trends and the oscillations of these curves, starting with absolute
admittance.

Absolute admittance (panels (a) and (b) of
figure~\ref{fig:Results_Case4_Admittance}) depends on the period of
sea-level oscillation, the permeability, and to a lesser extent, on
the half-spreading rate.  We consider these in turn.  Shorter periods
of sea-level variation at constant amplitude give larger values of
$\vert\dot{S}\vert$ and $\vert\dot{z}_m\vert$, and hence increase the
temporal variation of \fcotwo{}. This causes increased deviation of
ridge emissions from baseline. Increased mantle permeability and
spreading rate both reduce the melt travel-time from the base of the
melting region.  In the melt travel-time model of
equation~\eqref{eq:tau}, the permeability appears as $K_0$, while the
spreading rate $U_0$ is directly proportional to the mantle upwelling
rate $W_m$.  A reduced melt travel-time implies a smaller difference
between \taumax{} and \taumin{}, and therefore a focusing of \cotwo{}
from the base of the melting region to the ridge axis over a shorter
interval in time. The uncertainty in mantle permeability translates to
a broader spread of the absolute admittance curves than does the range
of half-spreading rates considered.

Relative admittance is equal to the absolute admittance normalised by
the baseline emissions rate.  The baseline depends on half-spreading
rate but not on permeability (fig.~\ref{fig:Results_Case1}(a)). We
therefore see a difference between absolute and relative admittance in
figure~\ref{fig:Results_Case4_Admittance}(b) and (d).  For
slow-spreading ridges, which have a low baseline emissions rate, the
normalised variance (and hence the relative admittance) is larger.

The oscillations superimposed on the primary admittance trend are not
physically significant, but are readily explained.  They arise from
variation in the number of sea-level half-cycles that fit into the
time interval from $t-\taumax$ to $t-\taumin$; this is the time
interval over which $\Sdot(t-\tau)$ can contribute \cotwo{} emissions
at time $t$.  For oscillatory sea level, each positive or negative
peak in $\Sdot$ has an opposing emissions effect relative to the prior
negative or positive peak. If there is an unmatched peak affecting the
bottom of the melting region, the amplitude of \ecotwo{} variations is
larger and the admittance is higher.

Broadly, the patterns of admittance and lag imply that, in terms of
\cotwo{} emission variation, the dominant sea-level changes will be
those with large amplitude and short period changes.  The emissions
variation associated with such changes will lag the forcing by
approximately $\taumin$.  The modelled magnitude and lag of \ecotwo{}
changes are affected by both $K_0$ and $U_0$.

\subsection{Reconstructed Pleistocene Sea Level}
\label{sec:reconstr-pleist-sea}
\begin{figure}[h!]
  \centering
  \includegraphics[width=16cm]{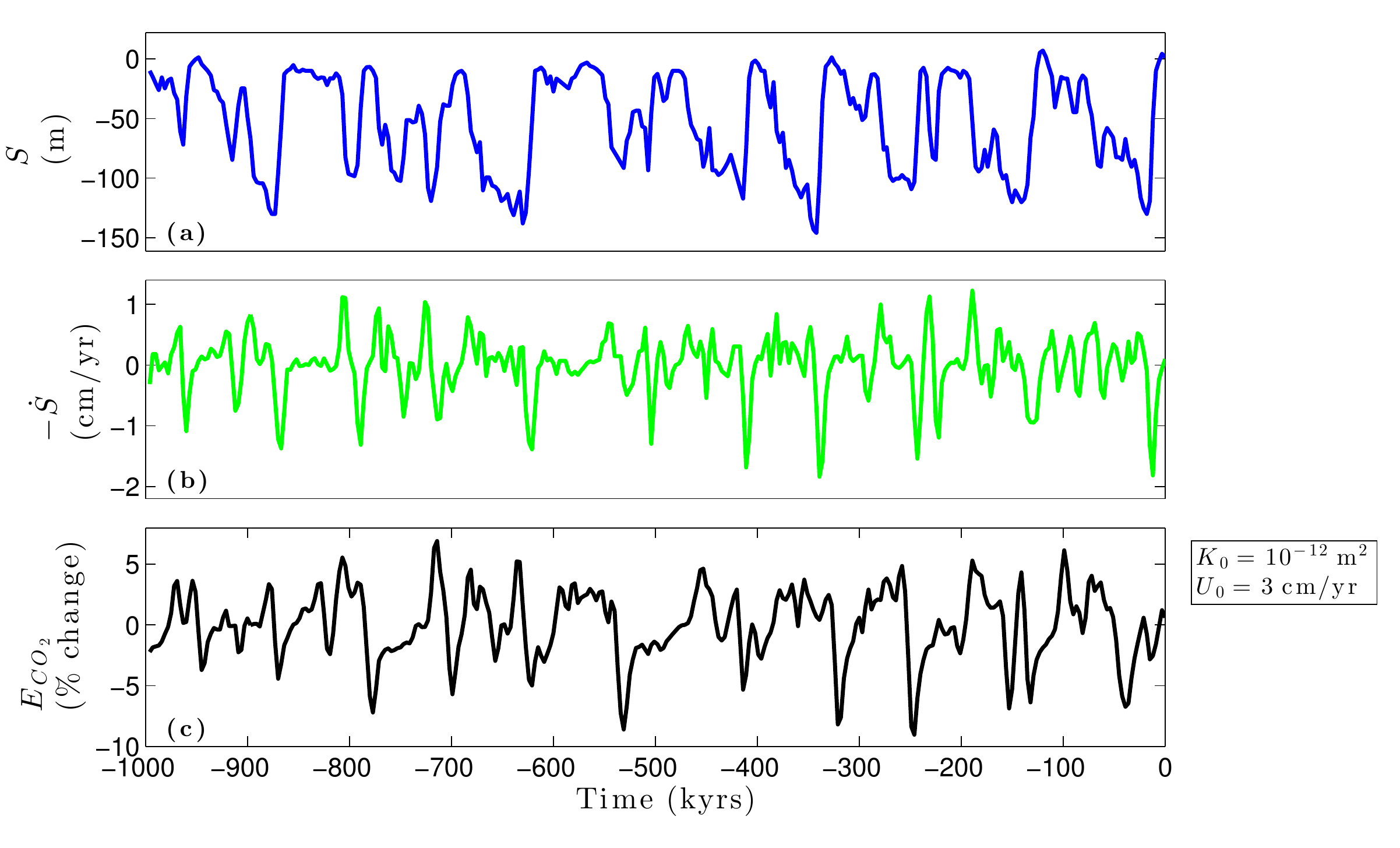}
  \caption{Reconstructed sea level. Plots show \textbf{(a)}
    sea-level change, \textbf{(b)} the negative rate of sea-level
    change, and \textbf{(c)} \cotwo{} emissions from a section of
    mid-ocean ridge.  The lag is $\sim$90~kyrs.}
  \label{fig:LocalEmissions_RealSL}
\end{figure}

The simple scenarios of sea-level variation presented above give
insight into the behaviour of the model, but are not representative of
the variations that have occurred naturally, over the past million
years.  We move, therefore, to a model forced by the time-series of
reconstructed global sea level from \cite{siddall10}, shown in
figure~\ref{fig:LocalEmissions_RealSL}(a). Other reconstructions
exist, but the differences between them are small enough that we
follow \cite{Crowley:2014vy} and consider only this one.
\cite{siddall10} record data every $3$~kyrs and, based on their
reconstruction, the highest rates of sea-level change
(fig.~\ref{fig:LocalEmissions_RealSL}(b)) meet the condition
$\textrm{max}\left(\dot{z}_m(t)\right)<W_m$ required for validity of
equation~\eqref{eq:CO2_flux_rate}.

Figure~\ref{fig:LocalEmissions_RealSL}(c) shows the result of applying
reconstructed sea level to ridge \ecotwo{}.  There is a $\pm 8\,\%$
range in \ecotwo{} for moderate half-spreading rate and permeability.
The \ecotwo{} curve is, qualitatively, an offset version of $\dot{S}$
with small variations smoothed out --- as expected from \ecotwo{}
being approximated by convolving \Sdot{} with the emissions response
in figure~\ref{fig:Results_Case2}(c).  Within this framework, we now
consider how to apply the model to global MOR emissions.

\subsection{Global Mid-Ocean Ridges}
\label{sec:the-global-model}
The global MOR system is composed of ridge segments spreading at
different rates, ranging from the ultra-slow Gakkel ridge to the fast
East Pacific Rise.  The baseline emissions depend on the
half-spreading rate, as does the character of the emissions response
to sea-level change.  The global response to sea-level change should
therefore be computed as the segment-scale response, integrated over
the global MOR system,
\begin{align}
  \gcotwo(t,K_0) =
  \int_0^{L_{\text{MOR}}} \ecotwo(t,K_0,U_0(l))\text{d}l,
  \label{eq:global_emit_continuous} 
\end{align}
where \gcotwo{} is global MOR emissions of \cotwo{} in kg/yr,
$l$ is arc length along the ridge, and $L_{\text{MOR}}$ is the total
length of the MOR system.  This integral can be approximated by
discretising the half-spreading rate into bins $U_{0i}$ and summing
the local response in each bin. A weighting is applied to each entry
in the sum to account for the total length of segments with
half-spreading rates in that bin. The sum is written as
\begin{align}
  \gcotwo(t,K_0) = \sum_{i=1}^{N} \ecotwo(t,K_0,U_{0i})\; L_i(U_{0i}),
  \label{eq:global_emit_sum}
\end{align}
where $N$ is the total number of spreading rate intervals to sum over
and $L_i(U_{0i})$ is the total length of MOR in a particular
spreading-rate bin.  The local emission rate \ecotwo{} in each bin is
computed by adopting the average half-spreading rate of the bin and
assuming that sea-level change is eustatic --- the same for all
segments globally.

\begin{figure}[h!]
  \centering
  \includegraphics[width=12cm]{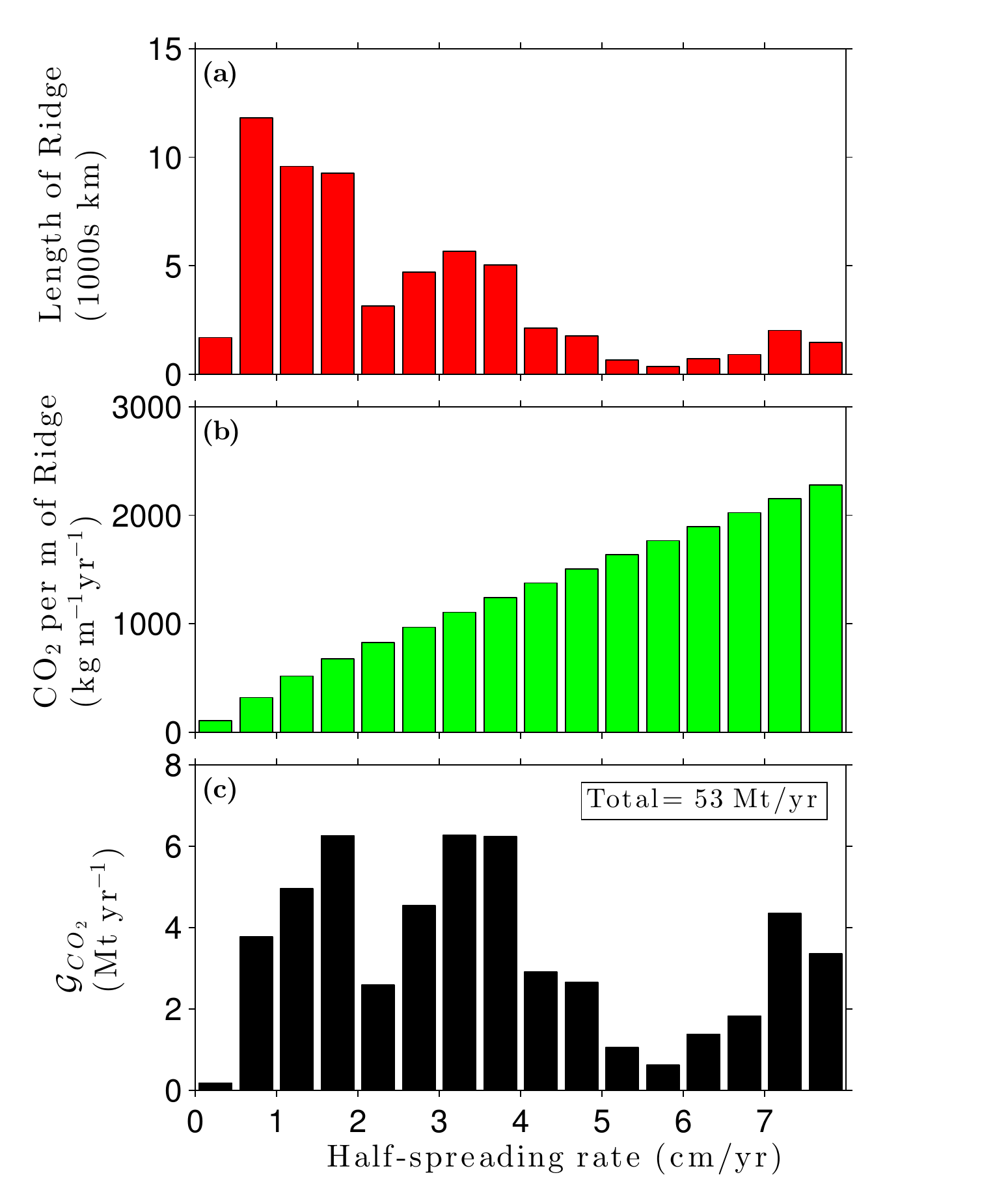}
  \caption{Quantities associated with discrete bins of half-spreading
    rate. \textbf{(a)} The total length of MOR
    \citep{Gale:2013fj}. \textbf{(b)} Baseline emissions per metre of
    MOR from our model. \textbf{(c)} Baseline global \cotwo{}
    emissions as the product of quantities in panels (a) and
    (b). Calculated for 0.38~kg \cotwo{} per m$^3$ of upwelling mantle
    (125~ppmw).  Baseline emissions are independent of $K_0$.}
  \label{fig:Global_CO2Histogram}
\end{figure}

\cite{Gale:2013fj} provide a catalogue of segment lengths and
spreading rates for the global MOR system; the total ridge length is
$61000$~km with a mean half-spreading rate of $2.5$~cm/yr.  A
histogram of these data is plotted in
figure~~\ref{fig:Global_CO2Histogram}(a).
Figure~\ref{fig:Global_CO2Histogram}(b) shows, for each spreading-rate
bin, the rate of baseline emissions per metre of ridge.  Panel (c)
then shows the product of ridge length and emissions rate per metre,
giving the total emissions rate associated with each spreading rate
bin.  These are summed in accordance with
eqn.~\eqref{eq:global_emit_sum} to give the total baseline global
response.  The global baseline emission rate thus predicted is 53~Mt
\cotwo{} per year assuming a sub-ridge mantle \cotwo{} concentration
of 125~ppmw.  This can be compared to other estimates of $91 \pm
45$~Mt\cotwo{}/yr \citep{Coltice:2004jl,Cartigny:2008dz,Marty:1998vo}.
If we instead constrain the model to have baseline emissions of
$91$~Mt\cotwo{}/yr, it requires a sub-ridge mantle \cotwo{}
concentration of $215$~ppmw (0.65~kg per m$^3$).

\begin{figure}[h!]
  \centering
  \includegraphics[width=16cm]{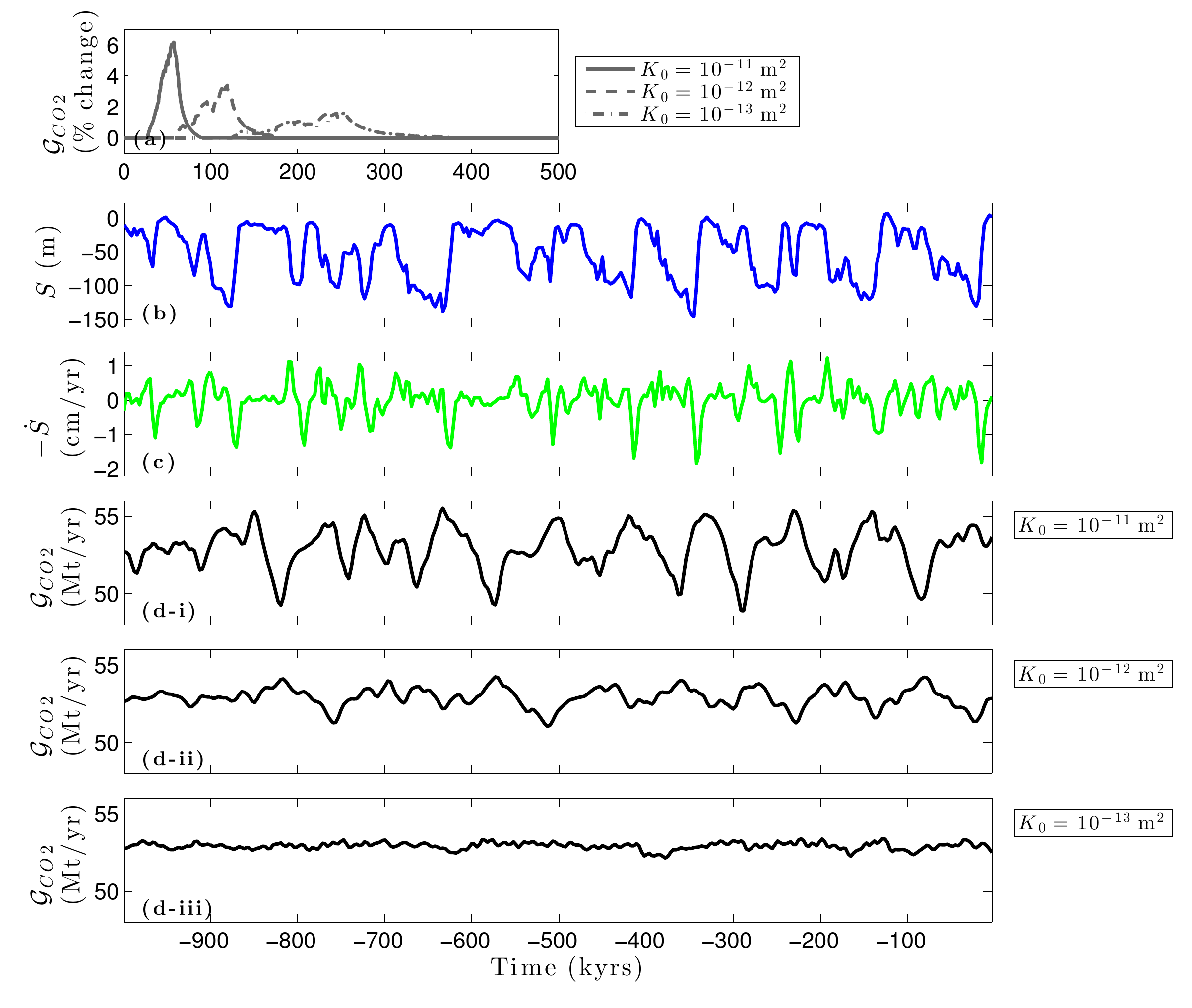}
  \caption{Global MOR emissions. Panel \textbf{(a)} shows \cotwo{}
    emissions from the global MOR system for a step change in
    sea-level identical to that of fig.~\ref{fig:Results_Case2}(a).
    Subsequent panels consider reconstructed sea level and its
    effect. \textbf{(b)} Reconstructed sea level; \textbf{(c)} the
    negative rate of sea-level change; and \textbf{(d-i--iii)}
    \cotwo{} emissions from the global MOR system.  The sea level
    time-series has been applied further back in time than shown, such
    that the left-most point in \gcotwo{} is affected by more than
    \taumax{}~kyrs of prior sea-level change. For (d-i,ii,iii) the
    lags in \gcotwo{} are, respectively, $\sim$60~kyrs,
    $\sim$120~kyrs, and $\sim$250~kyrs.}
  \label{fig:Global_Emissions_RealSL}
\end{figure}

Before applying the reconstructed sea-level forcing to the weighted
global emissions sum in eqn.~\eqref{eq:global_emit_sum}, we consider
the simple sea level forcing that was used to probe the behaviour of
\ecotwo{} in fig.~\ref{fig:Results_Case2}.  This linear ramp in sea
level is applied to compute the global emissions response in
figure~\ref{fig:Global_Emissions_RealSL}(a) for a range of mantle
permeabilities.  Global emissions in
figure~\ref{fig:Global_Emissions_RealSL}(a) are, unsurprisingly, more
complex than the \ecotwo{} equivalent, as they consist of a summation
of $N$ \ecotwo{} peaks from figure~\ref{fig:Results_Case2}, weighted
according to the ridge length in each bin and offset due to the
variation in $\tau$ with $U_{0i}$.  Compared with
figure~\ref{fig:Results_Case2}, \gcotwo{} has a smaller percentage
difference in the rate of \cotwo{} emissions and is spread out over a
longer time-period.

Finally, we apply the reconstructed sea-level time series to the
global model.  Figure~\ref{fig:Global_Emissions_RealSL} shows
\gcotwo{} for reconstructed sea level, calculated and plotted for a
set of three values of permeability constant $K_0$ in panels (d-i),
(d-ii), and (d-iii).  These curves demonstrate a reduction in
\gcotwo{} range from $6.6$~Mt\cotwo{}/yr for $K_0=10^{-11}$~m$^2$ to
$1.2$~Mt\cotwo{}/yr for $K_0=10^{-13}$~m$^2$.  The reduction in
\gcotwo{} range occurs because $K_0$ affects the range of $\taumin$
that is implicit in the global sum in
eqn.~\eqref{eq:global_emit_sum}. A large value of $K_0$ gives higher
overall permeability, shorter melt-transport times, and a global range
in \taumin{} that is smaller.  Therefore the emissions response to a
box-pulse in \Sdot{} (fig.~\ref{fig:Global_Emissions_RealSL}(a)) is
temporally concentrated and attains a higher peak value. Hence, larger
$K_0$ causes greater amplitude of variation in \gcotwo{} for
reconstructed sea level.

\section{Discussion}
\label{sec:discussion}
The model prediction for global MOR \cotwo{} emissions at constant sea
level using a mantle \cotwo{} concentration of 125~ppmw is $53$~Mt
\cotwo{} per year ($14$~Mt carbon per year), a value that is within
error of estimates from analyses of mid-ocean ridge basalts of $91 \pm
45$~Mt\cotwo{}/yr \citep{Coltice:2004jl,Cartigny:2008dz,Marty:1998vo}.
However, note that our prediction is not independent of MORB studies
because our choice of $\mathcal{C}_{\cotwo}$ is based on observed MORB
chemistry.  If we instead choose to constrain the model to fit
published \cotwo{} emissions, we require a source mantle \cotwo{}
concentration of $\sim$215~ppmw.

The ranges in global \cotwo{} emissions under reconstructed variation
of past sea level (fig.~\ref{fig:Global_Emissions_RealSL} for 125~ppmw
\cotwo{}) are $1$~Mt\cotwo{}/yr, $3$~Mt\cotwo{}/yr, and
$7$~Mt\cotwo{}/yr for mantle permeabilities at $1\%$ porosity of
$10^{-13}$~m$^2$, $10^{-12}$~m$^2$, and $10^{-11}$~m$^2$ respectively.
These idealised predictions assume that 100\% of \cotwo{} transported
to the ridge axis is degassed into the oceans, rather than retained in
the crust or mantle. This may, in fact, be rather accurate;
\cite{Cartigny:2008dz} estimated that over 80\% of \cotwo{} in
primitive MORB is degassed near the ridge axis.  Furthermore, 100\%
degassing is assumed in the papers that calculate the source mantle
\cotwo{} concentration we use.  It has previously been speculated
that variations in ridge \cotwo{} emissions would have an effect on
global climate \citep{Huybers:2009hn, Lund:2011jd}; it is beyond the
scope of the present manuscript to investigate this question, though
it is a target for future work.

Uncertainty in mantle permeability translates to uncertainty in both
the amplitude of variations and the lag of global \cotwo{} emissions
from MORs. There are various experimental constraints on mantle
permeability \citep[e.g.][]{Miller:2014gp, Connolly:2009jh}; these
tend to agree on the scaling with porosity, but disagree on the
magnitude of $K_0$.  Furthermore, permeability is sensitive to grain
size, a parameter that is poorly known for the mantle beneath MORs
\citep[although see][]{turner15}.  Our chosen range of $K_0$ is
intended to accommodate these uncertainties, as well as represent an
effective permeability for melt transport that may be channelised into
high-porosity, high-permeability dunite channels
\citep[e.g.][]{Jull:2002wz,Kelemen:1995hm}.  Our $K_0$ range of
$10^{-13}$--$10^{-11}$~m$^2$ encompasses a change in the amplitude of
\gcotwo{} variation by a factor of 5, a difference in lag of 200~kyrs,
and qualitative difference in the time-series of \gcotwo{}
(fig.~\ref{fig:Global_Emissions_RealSL}).  Therefore $K_0$ represents
a leading source of uncertainty in the model.  Uranium series
disequilibria may provide an independent constraint on magma travel
time from the base of the melting region
\citep[\textit{e.g.},][]{Jull:2002wz}, although interpreting the
various species in the decay chain is fraught with complexity.
Preservation of $^{230}$Th disequilibrium (half-life of $75$~kyrs) suggests a permeability of
$K_0 \ge 10^{-12}$~m$^2$, and community consensus similarly favours
$K_0$ at the higher end of our considered range.

Our model is based on the assumption that melt travels vertically from
the depth of first melting to the top of the melting region and is
then focused along a sloping, high porosity decompaction channel to
the ridge axis \citep{Sparks:1991th}.  The travel time of the vertical
flow is modelled by a 1D compaction column; we assume that transport
in the decompaction channel is instantaneous. The systematic error
introduced by the latter assumption is zero on the ridge axis ($x=0$),
and increases with distance $x$ from the ridge axis.  This means
$\tau$ plotted in figure~\ref{fig:Results_Case2_Tau}(b) is more
accurate at small $x$, but increasingly underestimates $\tau$ for
larger $x$.  Therefore, assuming $\taumin$ is accurately modelled,
$\taumax$ is probably too small, such that \ecotwo{} in
figure~\ref{fig:Results_Case2} should have a longer tail on the right
of the graph.  However, long tails have little effect on the resultant
\ecotwo{} pattern for complex sea-level changes or on the \gcotwo{}
pattern for reconstructed sea level.  Therefore the overall effect of
including a finite travel time along the high porosity channel would
be to make a small adjustment to the \ecotwo{} response.  This
suggests that assuming instantaneous travel time along the channel has
little effect on the results of the model.

Another assumption made is that travel time is constant with respect
to time, despite changes in melting rate (and thus, porosity) caused
by changing sea-level.  This follows the approach in the reduced model
of \cite{Crowley:2014vy}, where the perturbations in porosity were
taken as negligible disturbances to the travel time in a steady-state
compaction column solution from \cite{Hewitt:2010il}.

A more significant assumption underpinning the model is that carbon
dioxide behaves as a perfectly incompatible tracer --- meaning that
carbon concentration does not affect the mantle's physical or
thermodynamic properties.  However, carbon is not a trace element.  In
contrast, experiments by \cite{Kono:2014ja} document the very low
viscosity of insipient, carbon-rich melts present at small melt
fraction below the base of the silicate melting region.  The
experiments also show that viscosity rises sharply as the carbon is
diluted by silicate melting.  It would be challenging to capture this
variability in models, especially since the wetting properties (and
hence the mobility) of carbon-rich melts are poorly constrained.

A more significant concern, however, is that treating volatiles as
trace elements neglects their thermodynamic effect on melting.  Small
mantle concentrations of carbon affect the depth at which melting
begins, though the melt fractions produced by this incipient melting
are probably less than a few tenths of a percent
\citep{Dasgupta:2013us}.  Our model assumes that these melts do not
segregate until the onset of silicate melting.  At such small
porosity, it is unclear whether these carbonated melts can
percolate. However, water-induced melting at the wet peridotite
solidus of $\sim$90--120~km \citep{Asimow:2003dt, Dasgupta:2013us}
increases the melt fraction.  Again, the threshold of
interconnectivity for such melts is not known, so it is possible that
such deep, hydrous melts do not segregate, or do so very slowly. If
the $^{230}$Th disequilibrium observed in young MOR lavas originates
with melt segregation in the presence of garnet \citep{stracke06}, it
would support the hypothesis of efficient segregation of hydrous melts
(although other hypotheses also fit the observations).  Overall, our
model depends only on the presence of a pressure-dependent boundary
that separates non-segregating melts, below, from segregating melts,
above.  The sharpness that is required of this boundary is unclear.

Finally, our model assumes a chemically and thermally homogenous
mantle, which is certainly not true of the natural system
\citep[\textit{e.g.},][]{Dalton:2014fv}.  No data exists that would
allow us to accurately incorporate small-scale ($\lesssim 100$~km)
heterogeneity in model.  If such heterogeneity is pervasive and at
scales of $\sim$10~km or smaller, it would affect the style of melt
transport \citep{Katz:2012tz}, with fertile regions creating pathways
for rapid melt transport through the melting region.  It may be the
case that this is captured by a high effective permeability, though
this is probably not a testable hypothesis.  Large-scale heterogeneity
would leave the mantle homogenous over the scale at which we calculate
\ecotwo{}, so the underlying melt transport model would be unaffected,
though parameters would need to be adjusted according to the oceanic
region.  It seems likely that such variations would cancel in the
integral for global \cotwo{} emissions rate.

Previous authors have suggested that changing sea level might affect
MOR \cotwo{} emissions almost instantaneously
\citep{Huybers:2009hn,Lund:2011jd,Tolstoy:2015kq}.  Their assumption
is that pressure changes instantaneously affect melting rates, MOR
volcanic productivity and therefore, assuming constant \cotwo{}
concentration in the erupted melt, MOR \cotwo{} emissions.  We
disagree with the assumption that \cotwo{} concentrations would be
constant. \cotwo{} is highly incompatible and therefore additional
melting acts to dilute the constant mass of \cotwo{} in the
melt. However, after including these effects, our model can calculate
whether changing sea level affects MOR \cotwo{} emissions (see
appendix~\ref{sec:inst-co2-resp}).  To leading order there is no
effect; the reduced (or increased) concentration of \cotwo{} in the
melt counteracts the increased (or reduced) rate of melt delivery to
the ridge axis.

We would like to be able to compare the model to data, but there is no
dataset of global MOR \cotwo{} flux over time.  Atmospheric \cotwo{}
concentration from Antarctic ice cores \citep{Bereiter:2015js} is an
existing dataset that might record some influence of
\gcotwo{}. However, there are many strong, nonlinear controls on
atmospheric \cotwo{} and the relationship between \gcotwo{} and
atmospheric \cotwo{} will not necessarily be constant over time.
Nonetheless, comparing atmospheric \cotwo{} to \gcotwo{} does offer at
least one useful insight: if there is no correlation between these
quantities within the model's reasonable parameter space, it would be
an indication that MOR \cotwo{} emissions have no effect on
atmospheric \cotwo{}.  We find a correlation in
appendix~\ref{sec:corr-atmosph-co2}.  The best correlation, without
considering the very significant complicating factors associated with
climatic feedbacks, occurs at $K_0=10^{-11.37}$ with a lag of
$73$~kyrs.  This correlation does not prove any causative link between
\gcotwo{} and atmospheric \cotwo{}.

Variable MOR \cotwo{} emissions' effect on atmospheric \cotwo{} will
vary over time, as the fraction of MOR \cotwo{} emissions into the
intermediate ocean that reach the atmosphere is not constant.  This
fraction depends upon an array of factors that affect ocean
alkalinity, ocean upwelling patterns and \cotwo{} concentration in the
atmosphere, and these factors vary on sub-glacial/glacial timescales
(as do numerous processes interacting with them, \textit{e.g.,} the
biosphere) \citep{BROECKER:1982ex, WALKER:1986ki,
  Brovkin:2012ir, Yu:2014vi}.  This would have to be taken into
account if inferring any climate implications from our model.

The MOR carbon flux may vary over time in other ways that we have not
considered here.  For instance, it is plausible that the intensity of
hydrothermal circulation varies with sea level, driven by variations
in melt supply \citep{Lund:2011jd, Crowley:2014vy}.  If this is the
case, hydrothermal variations would have a different lag than that of
\cotwo{} emissions. Hydrothermal systems have been proposed as both a
\cotwo{} sink, with hot seawater transforming basalts to clay
\citep{Gysi:2012hb}, and a \cotwo{} source, with hydrothermal fluids
transporting \cotwo{} from magma to the ocean \citep{Sakai:1990vw}.
The rate of both these processes might scale with hydrothermal
circulation, although it is not clear whether the net effect would be
to increase or decrease MOR \cotwo{} emissions.

\section{Summary} 
\label{sec:conclusion}
The model presented above builds on the reduced model of
\cite{Crowley:2014vy} to calculate the efflux of a highly incompatible
chemical component from a mid-ocean ridge, and how that efflux would
vary with changes in sea level. It is based on a description of melt
transport through a homogenous mantle and assumes perfect
incompatibility of the component.  This leads to a simple but
physically consistent model of chemical transport through the melting
region beneath a ridge.  The model calculates total melt supply rate
and global background emissions of \cotwo{} that are consistent with
data and prior estimates.

In the model, changing sea level affects the depth of first silicate
melting; this alters the rate at which \cotwo{} enters the melting
region, segregates from its mantle source, and (some time later)
arrives at the ridge axis and is degassed into the ocean.

The MOR emission rate of \cotwo{} is predicted to vary by up to
$12.5\%$ when the model is forced with reconstructed Pleistocene sea
level variation ($7$~Mt\cotwo{}/yr for 125~ppmw
$\mathcal{C}_{\cotwo}$; $11$~Mt\cotwo{}/yr for 215~ppmw
$\mathcal{C}_{\cotwo}$).  There is uncertainty in the predicted
magnitude and timing (relative to sea-level forcing) of this effect,
as two parameters of the model --- $\mathcal{C}_{\cotwo}$ and $K_0$
--- are weakly constrained by existing data.  However, within
reasonable ranges of the model parameters, the amplitude of global MOR
\cotwo{} emission-rate variation will remain on the order of several
Mt\cotwo{}/yr. The total difference in the mass of \cotwo{} emitted
from MORs during sea-level driven deviations from global baseline
\cotwo{} is up to $\sim$80~Gt \cotwo{} for high permeability and
125~ppmw $\mathcal{C}_{\cotwo}$ (see
figure~\ref{fig:Global_Emissions_RealSL}(d-i)).  This is $4\%$ of the
pre-industrial \cotwo{} mass in the atmosphere of $2190$~Gt, or
$0.06\%$ of pre-industrial \cotwo{} in the oceans \citep[IPCC:][fig
7.3]{Solomon:2012uy}.

Our results indicate that the \cotwo{} emissions from mid-ocean ridges
are temporally variable in response to sea-level change.  These
results align with the hypothesis of \cite{Huybers:2009hn}; however,
whereas they assumed an immediate emissions response, we show that the
mid-ocean ridge \cotwo{} emissions response lags the sea-level forcing
by approximately the minimum travel time of \cotwo{} through the
melting region --- at least $60$~kyrs. Therefore MOR \cotwo{}
emissions cannot feed back into the sea-level change that caused them;
instead these \cotwo{} emissions will enter the climate system during
the next glacial cycle.  Further work is needed to assess the climatic
impact of the variable \cotwo{} emissions predicted for mid-ocean
ridges in this paper and for subaerial volcanoes by
\cite{Huybers:2009hn}.

\paragraph{Acknowledgements} The research leading to these results has
received funding from the European Research Council under the European
Union's Seventh Framework Programme (FP7/2007-2013) / ERC grant
agreement number 279925. Thanks to P.~Huybers, C.~Langmuir,
P.~Kelemen, D.~Schrag, and members of the FoaLab for helpful
discussions, and to C.~Rowan for a spreading-rate dataset that was
used to test the model. The paper was improved on the basis of helpful
reviews by A.~Stracke and two anonymous reveiwers.  Katz thanks the
Leverhulme Trust for additional support.

\newpage{}
\appendix
\section{Appendix}
\label{sec:appendix}

\subsection{Depth of First Silicate Melting}
\label{sec:deriv-depth-melt}

The depth of first silicate melting occurs at the intersection of the
local mantle temperature, which is adiabatic prior to first melting,
and solidus temperature.  We take a standard adiabatic temperature
expression and approximate it with a Taylor series truncated at first
order in pressure to obtain
\begin{equation}
  \label{eq:AdiabatTemp}
  T(z) = \widetilde{T}-\frac{\alpha g \widetilde T}{\rho c}
  \left(\rho z -\rho_w S\right) \;\;, 
\end{equation}
where symbols are as defined in the main text and we have added a
pressure term for varying sea level.  The mass of a column of water is
$\rho_w S$, provided the thermal expansion of the ocean is much less
than $S$.  Fortunately, oceanic thermal expansion causes less than
$1\%$ of total glacial sea-level change
\citep{McKay:2011gs}. Freshwater density is used for $\rho_w$, as salt
remains in the ocean when water is removed, and ocean water inputs are
fresh. Coordinates are upwards-positive with origin at the MOR,
therefore $z$ is negative and increasing sea level $S$ is positive.
We model the solidus temperature following \cite{Katz:2008gd} and
\cite{Hewitt:2010il}, using a solidus that is linear in pressure and a
single composition parameter.  We consider pressure terms due to the
mass of mantle above the solidus, and sea level deviations from
reference conditions.  This gives
\begin{equation}
  T_{\textrm{Solidus}} = T_{S_{\textrm{ref}}} - 
  \gamma g \left(\rho z -\rho_w S\right)\;\;. 
  \label{eq:SolidusTemp}
\end{equation}
We ignore bulk ocean+atmospheric weight as this is accounted for in
$T_{S_{\textrm{ref}}}$. Setting $T_{\textrm{Solidus}}$ in
eqn.~\eqref{eq:SolidusTemp} equal to $T(z)$ in
eqn.~\eqref{eq:AdiabatTemp} and solving for depth we obtain
eqn.~\eqref{eq:MeltingDepth} in the main text.

\subsection{ Upper Boundary of the Melting Region}
\label{sec:deriv-upper-bound}

The upper boundary of the melting region, like the lower boundary, is
a region where the solidus temperature matches the local mantle
temperature.  Local mantle temperature in the melting region can be
modelled as the adiabatic temperature profile
(eqn.~\eqref{eq:AdiabatTemp}) plus terms for \textit{(i)} thermal
energy lost to latent heat during melting of $(z-z_m)\Pi L/\rho c$,
and \textit{(ii)} conduction near the cold ceiling (\textit{i.e.},\ the
ocean floor) from the half-space cooling model.  This approach assumes
superposition of temperature fields without cross terms correcting for
deviations from the assumptions of each model (\textit{e.g.},
half-space cooling assumes a constant background temperature with
respect to depth).  Matching this to solidus temperature from
equation~\eqref{eq:SolidusTemp} gives
\begin{align}
  T_{S_{\textrm{ref}}}- \gamma \rho gz 
                      &= \widetilde{T} -\frac{\alpha g \widetilde{T}}{c}z
                        -\frac{\Pi L}{\rho c}(z-z_m)
                        - (\widetilde{T}-T_{0})\,\text{erfc}
                        \left(\frac{z}{2}\sqrt{\frac{U_0}{\kappa
                        x}}\right)\;\;,
   \label{eq:UpperMeltBoundary_FullEquation}
\end{align}
where $L$ is the latent heat capacity of the mantle.

Equation~\eqref{eq:UpperMeltBoundary_FullEquation} cannot be
rearranged to give $z$ in terms of $x$.  However we can get $x$ in
terms of $z$, as shown in equation~\eqref{eq:x_l}, an expression shown
to accurately match melting region from numerical studies in
\cite{Crowley:2014vy}. This justifies the superposition of temperature
fields assumption and gives us a reference to benchmark against,
\begin{equation}
  x_l(z) = \frac{U_0 z^2}{4 \kappa} \left( \textrm{erfc}^{-1} 
    \left[\frac{\gamma \rho g-\alpha g \widetilde{T}/c-\Pi
        L/\rho c}{\widetilde{T}-T_{0}} 
      \left(z-z_m\right) \right] \right)^{-2} \;\;.
  \label{eq:x_l}
\end{equation}

\begin{figure}[ht]
  \centering
  \includegraphics[width=8cm]{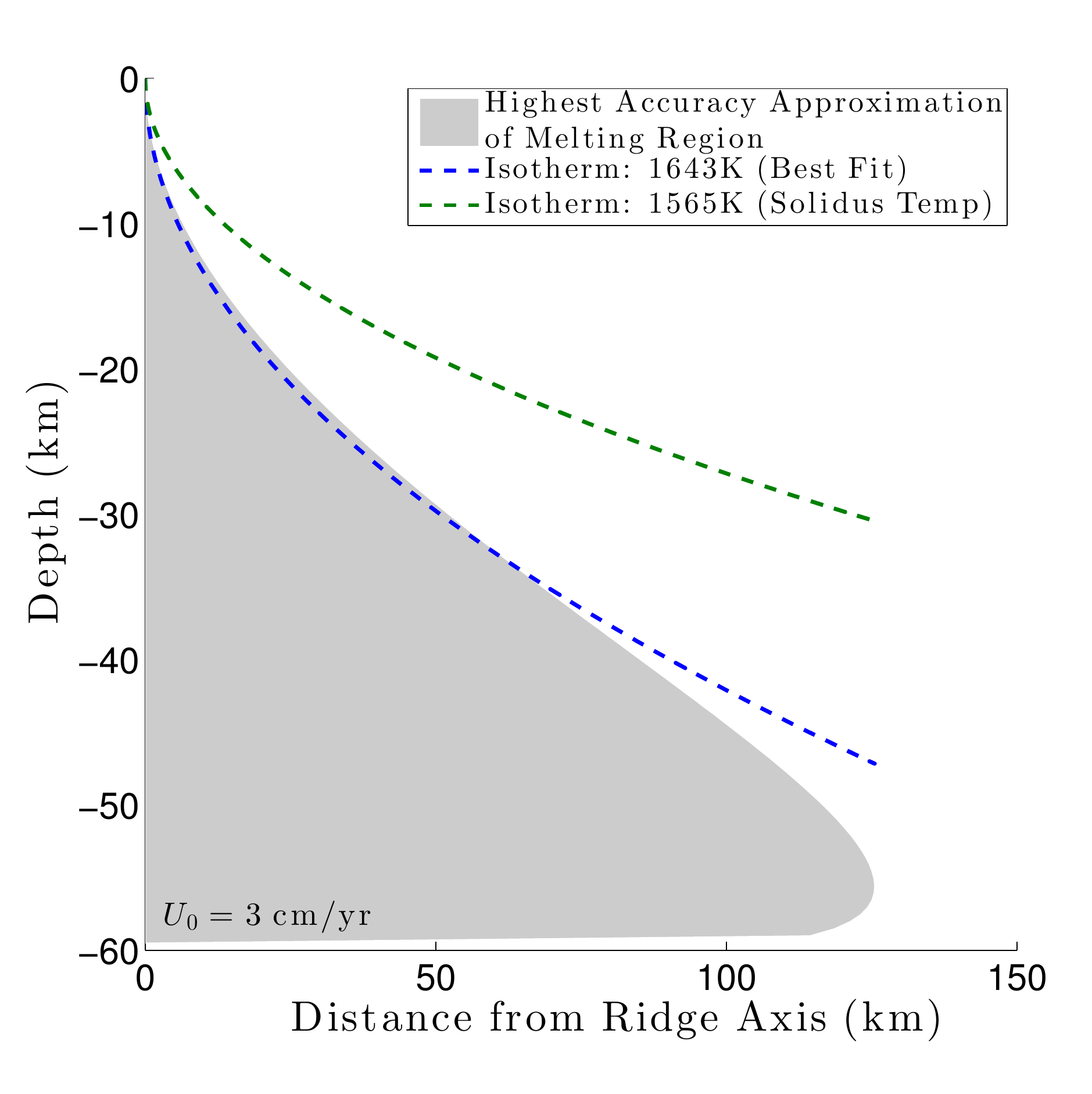}
  \caption{Melting-region calculation from \citep{Crowley:2014vy},
    compared to isotherms of the half-space cooling solution for
    1565~K (a realistic solidus temperature) and 1643~K (a best fit
    isotherm). The 1643~K isotherm is used as the upper boundary of
    the melting region for this paper. All curves are calculated for a
    mantle potential temperature of 1648~K, surface temperature of
    273~K and a half-spreading rate of 3~cm/yr.  Plotted quantities
    all scale similarly with $U_0$ and thus isotherms have the same
    relative error across all spreading rates.}
  \label{fig:ApproximatingMeltRegionWith_HSC_Model}
\end{figure}

To get $z$ in terms of $x$, we discard all depth-dependent temperature
terms in equation~\eqref{eq:UpperMeltBoundary_FullEquation} except
half-space cooling.  This approach is equivalent to plotting an
isotherm in a half space cooling model. Thus
\begin{align}
  z_l(x,U_0)= 2\sqrt{\frac{\kappa x}{U_0}}\; \text{erfc}^{-1}
  \left(\frac{\widetilde{T}-T_l}{\widetilde{T}-T_0}\right)\;.
  \label{eq:Lithospheric_thickness}
\end{align}
Figure~\ref{fig:ApproximatingMeltRegionWith_HSC_Model} compares
equation~\eqref{eq:Lithospheric_thickness} to the benchmark
(eqn.~\eqref{eq:x_l}).  For a lithospheric boundary temperature of
$\sim$1550~K --- close to the solidus temperature of mantle ---
equation~\eqref{eq:Lithospheric_thickness} is a poor approximation of
the real melting region. However, unphysically high boundary
temperatures of $\gtrsim 1640$~K give good approximations of the
melting region, as this compensates for the temperature effects not
explicitly accounted for.  Therefore, we use
equation~\eqref{eq:Lithospheric_thickness} with $T_{l}=1643$~K to
approximate the upper boundary of the melting region.

\subsection{Width of the Melting Region} 
\label{sec:deriv-melt-width}

Equation~\eqref{eq:x_l} shows a finite width of the melting region
defined by $\textrm{d}x_l / \textrm{d}z=0$ (see
figure~\ref{fig:ApproximatingMeltRegionWith_HSC_Model} at
$z\simeq55$~km, $x\simeq 130$~km).  The derivative of
equation~\eqref{eq:x_l} is
\begin{align}
  \diff{x_l(z)}{z} = \frac{U_0 z}{4 \kappa} 
  &\left(\frac{2\,\textrm{erfc}^{-1}\big(B(z-z_m)\big)  -  
    \sqrt{\pi} B \, \textrm{e}^{\left[\textrm{erfc}^{-1} (B(z-z_m))
    \right]^2}}{\left[\textrm{erfc}^{-1}
    \big(B(z-z_m)\big)\right]^3}\right) \;\;,
    \label{eq:deriv_xl_wrt_z} \\  
  & \;\;\;\;\textrm{where:}\;\;\; 
    B=\frac{\gamma \rho g-\alpha g 
    \widetilde{T}/c-\Pi L/\rho c}{\widetilde{T}-T_{0}}\;\;.
\end{align}

We have not found an analytical solution to
equation~\eqref{eq:deriv_xl_wrt_z} for
$\textrm{d}x_l / \textrm{d}z=0$.  However, computational investigation
of equation~\eqref{eq:x_l} for varying $U_0$ shows that the depth of
the widest point of the melting region is constant. Calculating $x_l$
at this fixed depth gives the maximum width of the melting region.
Thus,
\begin{align}
  \textrm{max}(x_l) = x_w =  \frac{U_0 z_w^2}{4 \kappa} 
  \left( \textrm{erfc}^{-1} \left[ B \left(z_w-z_m\right) \right]
  \right)^{-2} \;\;,
  \label{eq:x_l_max_wrt_U0} 
\end{align}
where $z_w$ is the depth of the widest point of the melting region.

\subsection{MOR CO2 Emissions Response to Changing Melting Rate} 
\label{sec:inst-co2-resp}

We wish to know the effect of changing melting rates (due to pressure
changes) on the MOR efflux of \cotwo{}.  Following the model used in
the main text, the \cotwo{} emissions rate from a one-dimensional
column in the melting region is:
\begin{equation}
  \ecotwo  = R \mathcal{C}_{\cotwo}^{\textrm{melt}}\;\;,
  \label{eq:E_Rconc}
\end{equation}
where $R$ is the rate at which melt upwells per unit cross-sectional
area (kg~m$^{-2}$~s$^{-1}$), and
$\mathcal{C}_{\cotwo}^{\textrm{melt}}$ is the mass concentration of
\cotwo{} in the melt, both defined at the top of the column.  The
right hand side of eqn~\eqref{eq:E_Rconc} is affected by changing
pressure and its consequent effects on melting rate.

Increased melting rate has two effects.  Firstly, it increases
porosity, which induces faster melt upwelling (increasing $R$).
Secondly, it dilutes \cotwo{} in the melt, reducing
$\mathcal{C}_{\cotwo}^{\textrm{melt}}$ (\cotwo{} is highly
incompatible, thus its mass in the melt stays constant regardless of
changes to the total melt mass).  These two effects act in opposition,
and the net result on \ecotwo{} is not immediately obvious.  However,
it can be evaluated by calculating the rate of change of \ecotwo{} in
response to changing pressure,
\begin{equation}
  \diff{\ecotwo}{t} = \diff{R}{t} \mathcal{C}_{\cotwo}^{\textrm{melt}}  
  + \diff{\mathcal{C}_{\cotwo}^{\textrm{melt}}}{t} R
  \;\;. \label{eq:Edot_Raw}
\end{equation}
This calculation requires expressions for
$\mathcal{C}_{\cotwo}^{\textrm{melt}}$, $R$ and their derivatives.
Taking \cotwo{} as perfectly incompatible,
\begin{subequations}
  \begin{align}
    \mathcal{C}_{\cotwo}^{\textrm{melt}}  
    &= \frac{\mathcal{C}_{\cotwo}}{F}  \;\;, \label{eq:C}\\
    \diff{\mathcal{C}_{\cotwo}^{\textrm{melt}}}{t}  
    &= -\frac{\mathcal{C}_{\cotwo}}{F^2} \diff{F}{t}  \;\;,\label{eq:C_dot}
  \end{align}
\end{subequations}
where $F$ is the degree of melting. For a linearised solidus,
following \cite{Katz:2008gd} \cite{Hewitt:2010il}, this is
\begin{subequations}
  \begin{align}
    F &= \frac{\Pi}{\rho^2 g}
        \left( P - P_m \right)\;\;, \label{eq:F}\\ 
    \diff{F}{t} &= \frac{\Pi \rho_w}{\rho^2 g}\diff{S}{t}\;\;, \label{eq:F_dot}
  \end{align}
\end{subequations}
where $P$ is pressure, $P_m$ is the pressure at first silicate melting
and we have used $\dot{P}=\rho_w \dot{S}$.

$R$ is the product of the local porosity $\phi$ and melt upwelling
velocity $w$.  For a compaction column, following
\cite{Hewitt:2010il}, these are
\begin{subequations}
  \begin{align}
    w &=
        \left(\frac{K_0\Delta\rho g}{\eta_f}
        \right)^{\frac{1}{n}}
        \left(\frac{\Pi W_m}{\rho^2}
        \right)^{1-\frac{1}{n}} 
        \left(P-P_m\right)^{1-\frac{1}{n}} \;\;,\\
    \phi &=
           \left(\frac{K_0\Delta\rho g}{\eta_f}
           \right)^{-\frac{1}{n}}
           \left(\frac{\Pi W_m}{\rho^2}
           \right)^{\frac{1}{n}} 
           \left(P-P_m\right)^{\frac{1}{n}} \;\;.
  \end{align}
\end{subequations}
Thus $R=\phi w $ is
\begin{subequations}
  \begin{align}
    R &=  \left(\frac{\Pi W_m}{\rho^2}\right)
        \left(P-P_m\right)  \;\;, \label{eq:R}\\
    \diff{R}{t} &=  \left(\frac{\Pi W_m \rho_w}{\rho^2}\right)
                  \diff{S}{t}  \;\;. \label{eq:R_dot}
  \end{align}
\end{subequations}  

Substituting
equations~\eqref{eq:C}, \eqref{eq:C_dot}, \eqref{eq:F_dot}, \eqref{eq:R}, and
\eqref{eq:R_dot} into equation \eqref{eq:Edot_Raw} then rearranging gives
\begin{align}
  \diff{\ecotwo}{t} &= \left(1-\frac{\Pi (P-P_m)}{\rho^2 g F}\right)
                      \frac{\mathcal{C}_{\cotwo} \Pi W_m \rho_w}{\rho^2 F}
                      \diff{S}{t} = 0  \;\;,\label{eq:E_dot_rearrange_FINAL}
\end{align}
where, following eqn~\eqref{eq:F}, the term inside parentheses is
$1-F/F = 0$.  Therefore, to leading order, the instantaneous effect of
changing sea-level on MOR \cotwo{} emissions (through changes to
melting rate and eruption volume) is zero.  Thus, suggestions that MOR
\cotwo{} emissions should scale with eruptive volume are likely
incorrect; it is only more compatible elements that increase emissions
rates with increased melting.

However, the argument presented above is a simplification.  For
instance it assumes all melt that reaches the ridge is erupted
instantaneously. A more detailed model might consider magma reservoirs
at the MOR which could be vented by abrupt changes in pressure.
However, such ventings would only be fluctuations against the
background supply of \cotwo{} to the ridge, which (as shown above)
does not vary with melting rate.

\subsection{Correlation to Atmospheric CO2} 
\label{sec:corr-atmosph-co2}
Whilst there is no dataset of global MOR \cotwo{} flux over time, we
can compare the model's calculated MOR \cotwo{} flux to the related
quantity of atmospheric \cotwo{}.  Specifically, we can use
atmospheric \cotwo{} concentrations from amalgamated Antarctic ice
core data \citep{Bereiter:2015js}.  We do not believe there will be a
simple correlation between global MOR carbon emissions \gcotwo{} and
atmospheric \cotwo{}, nor do we think that this carbon flux is large
enough to dominate the atmospheric \cotwo{} variations.  However, if
there is no correlation between these quantities within the model's
reasonable parameter space, it would be a strong indication that MOR
\cotwo{} emissions have no effect on atmospheric \cotwo{}.

\begin{figure}[ht]
  \centering
  \includegraphics[width=17cm]{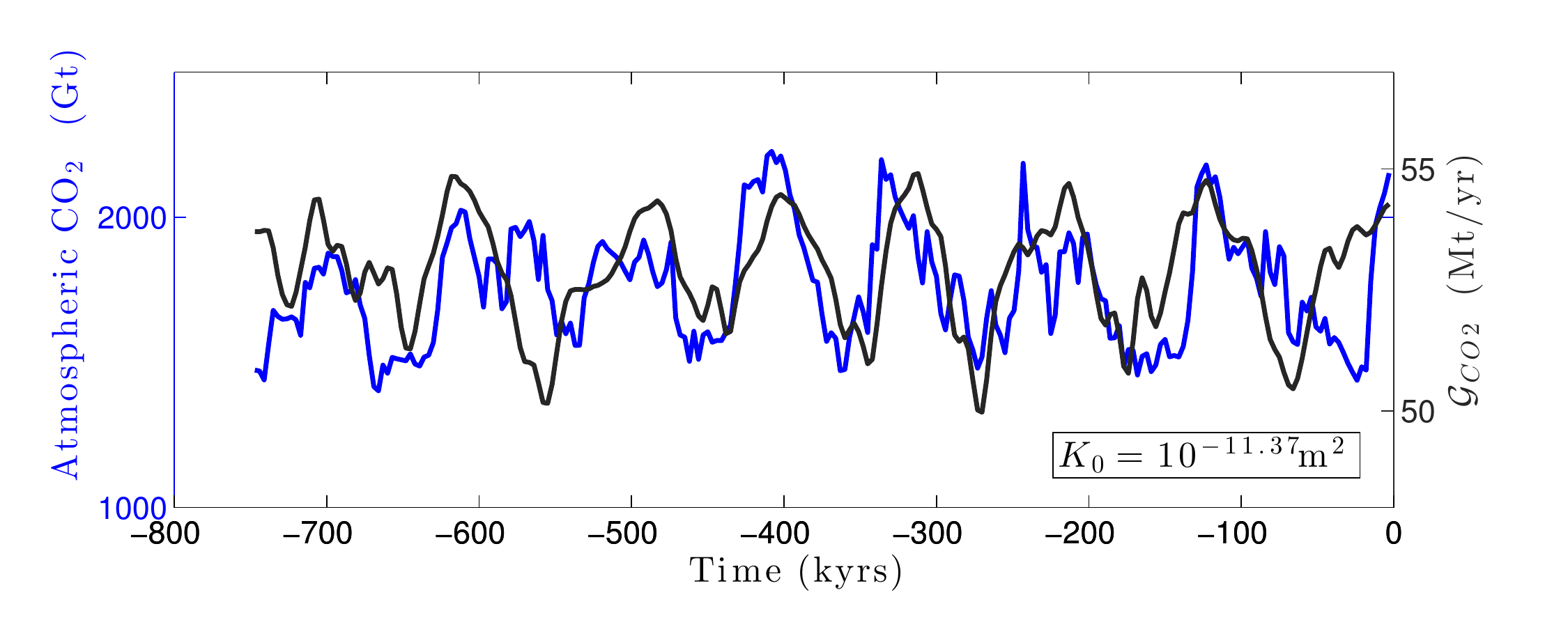}
  \caption{Best fit of global MOR \cotwo{} emissions (125~ppmw
    $\mathcal{C}_{\cotwo}$) to atmospheric \cotwo{} mass.  Atmospheric
    \cotwo{} mass was calculated using 1 ppmv \cotwo{} = 7.81~Gt
    \cotwo{}. Atmospheric \cotwo{} mass has been plotted to
    demonstrate that \gcotwo{} variations of a few Mt/yr
    ($10^{9}$~kg/yr) for 10's of kyrs ($10^{4}$~yr) have less mass
    than the change in atmospheric \cotwo{} mass ($10^{14}$~kg) by
    1--2 orders of magnitude. Therefore correlation between these
    quantities would have to be the result of feedbacks on atmospheric
    \cotwo{}.  Best fit was found by varying $K_0$ from $10^{-13}$ to
    $10^{-11}$ and finding the largest Pearson product-moment
    correlation coefficient (0.45).  Maximising Spearman (0.48) or
    Kendall (0.32) correlations produces the same $K_0$ value.  The
    lag for best fit $K_0$ of $10^{-11.37}$~m$^2$ is 73~kyrs.}
  \label{fig:BestFit2Data}
\end{figure}

In plotting atmospheric \cotwo{} (mass) against \gcotwo{} (mass/yr),
figure~\ref{fig:BestFit2Data} assumes that the atmosphere
instantaneously adjusts its equilibrium \cotwo{} concentration in
response to any change in outgassing at MORs.  This is a reasonable
timescale approximation, as the equilibrium timescales are short
compared to our study. However, assuming a linear change in
equilibrium atmospheric \cotwo{} concentration in response to
\gcotwo{}, despite atmospheric \cotwo{} having multiple feedbacks and
the time-varying fraction of MOR emissions entering the atmosphere
(see section~\ref{sec:discussion}), is a considerable flaw.

Figure~\ref{fig:BestFit2Data} shows that a correlation between these
series is possible within the reasonable parameter space of our
model's least-well constrained parameter $K_0$, however this is not a
proof of causal links. The process we have used could even be
considered circular. Atmospheric \cotwo{} and sea level are strongly
correlated and our model is driven by (rate of change of) sea level;
it is perhaps unsurprising that the model output is also correlated
with atmospheric \cotwo{}.

\newpage{}
\bibliographystyle{abbrvnat} 
\bibliography{manuscript}

\end{document}